\documentclass[]{spie}  
\usepackage[]{graphicx}
\usepackage{amssymb, amsmath}

\newcommand{\lya}{Ly$\alpha$}
\newcommand{\ha}{H$\alpha$}
\newcommand{\hb}{H$\beta$}
\newcommand{\oii}{[O II]}
\newcommand{\oiii}{[O III]}
\newcommand{\nii}{[N II]}
\newcommand{\heii}{HeII}
\newcommand{\arcmin}{$^{\prime}$}
\newcommand{\arcsec}{$^{\prime\prime}$}
\newcommand{\degree}{$^{\circ}$}

\title{Design and construction progress of LRS2-B: a new low resolution integral field spectrograph for the Hobby-Eberly Telescope\footnote{$\;$The Hobby-Eberly Telescope is operated by McDonald Observatory on behalf of the University of Texas at Austin, the Pennsylvania State University, Ludwig-Maximillians-Universit\"{a}t M\"{u}nchen, and Georg-August-Universit\"{a}t Goettingen.}} 

\author{Taylor S. Chonis\supit{a}, Hanshin Lee\supit{b}, Gary J. Hill\supit{b}, Mark E. Cornell\supit{b}, Sarah E. Tuttle\supit{b}, Brian L. Vattiat\supit{b}
\skiplinehalf
\supit{a}The University of Texas at Austin, Department of Astronomy, 2515 Speedway, Stop C1400, Austin, TX, USA 78712; \\
\supit{b}The University of Texas at Austin, McDonald Observatory, 2515 Speedway, Stop C1402, Austin, TX, USA 78712; \\
}

\authorinfo{Further author information: (Send correspondence to T.S.C.)\\T.S.C.: E-mail: tschonis@astro.as.utexas.edu, Telephone: 1-512-471-1495}

\begin{document} 
  \maketitle 

\begin{abstract}
\textbf{Note from the author (July 25, 2014): this paper has been superseded by Ref. \citenum{Chonis14} (arXiv:1407.6016)}.\newline

The upcoming Wide-Field Upgrade (WFU) has ushered in a new era of instrumentation for the Hobby-Eberly Telescope (HET). Here, we present the design, construction progress, and lab tests completed to date of the blue-optimized second generation Low Resolution Spectrograph (LRS2-B). LRS2-B is a dual-channel, fiber fed instrument that is based on the design of the Visible Integral Field Replicable Unit Spectrograph (VIRUS), which is the new flagship instrument for carrying out the HET Dark Energy eXperiment (HETDEX). LRS2-B utilizes a microlens-coupled integral field unit (IFU) that covers a 7\arcsec$\times$12\arcsec\ area on the sky having unity fill-factor with $\sim$300 spatial elements that subsample the median HET image quality. The fiber feed assembly includes an optimized dichroic beam splitter that allows LRS2-B to simultaneously observe $370 < \lambda \mathrm{(nm)} < 470$ and $460 < \lambda \mathrm{(nm)} < 700$ at fixed resolving powers of $R \approx \lambda/\Delta\lambda \approx 1900$ and 1200, respectively. We discuss the departures from the nominal VIRUS design, which includes the IFU, fiber feed, camera correcting optics, and volume phase holographic grisms. Additionally, the motivation for the selection of the wavelength coverage and spectral resolution of the two channels is briefly discussed. One such motivation is the follow-up study of spectrally and (or) spatially resolved \lya\ emission from $z \approx 2.5$ star-forming galaxies in the HETDEX survey. LRS2-B is planned to be a commissioning instrument for the HET WFU and should be on-sky during quarter 4 of 2013. Finally, we mention the current state of LRS2-R, the red optimized sister instrument of LRS2-B.
\end{abstract}


\section{INTRODUCTION} \label{sec:intro}  
The 10 m Hobby-Eberly Telescope (HET) is currently in the process of undergoing a major wide-field upgrade (WFU; Ref. \citenum{Hill12b}) that will increase its field of view to 22\arcmin\ in diameter and feature improved performance in preparation for the upcoming HET Dark Energy eXperiment (HETDEX; Ref. \citenum{Hill08a}). HETDEX will amass a sample of $\sim$0.8 million \lya\ emitting galaxies (LAE) to be used as tracers of large-scale structure for constraining dark energy and measuring its possible evolution from $1.9 < z < 3.5$. To carry out the 120 night blind spectroscopic survey covering a 420 square degree field (9 Gpc$^{3}$), the HET will be outfitted with a revolutionary new multiplexed instrument called the Visible Integral Field Replicable Unit Spectrograph (VIRUS; Ref. \citenum{Hill12a}). VIRUS consists of at least 150 copies (with a goal of 192) of a simple fiber-fed integral field spectrograph and for the first time has introduced industrial-scale replication to optical astronomical instrumentation. VIRUS is one of several instruments being prepared for the new era of HET instrumentation that has been ushered in by the WFU. For the past decade, the current Marcario Low Resolution Spectrograph (LRS; Ref. \citenum{Hill98}) has been a workhorse instrument for the HET, but its design is incompatible with the WFU. This provides an opportunity to redesign and improve upon the capabilities of LRS in a second generation instrument. 

The design of the second generation LRS (LRS2) is based on the versatile VIRUS unit spectrograph, which was designed to be adaptable to a range of spectral resolutions and wavelength coverage configurations. The original LRS2 design concept (see Ref. \citenum{Lee10}) is the first demonstration of the wide range of applications for the basic VIRUS design. The original LRS2 concept is fed by a 7\arcsec$\times$12\arcsec\ unity fill-factor integral field unit (IFU) and simultaneously covers $350 < \lambda (\mathrm{nm}) < 1100$ at a fixed resolving power of $R = \lambda / \Delta\lambda \approx 1800$. The wide spectral coverage is made possible by utilizing two VIRUS unit pairs (i.e., four total spectrograph channels). Utilizing the VIRUS unit spectrograph as the building block for LRS2 has allowed us to take advantage of the large engineering investment made in VIRUS for optimizing it for mass production. LRS2 will be built on the VIRUS production line, which will greatly reduce the final cost and delivery time to relatively low levels for such a capable instrument. While much of the design concept discussed in Ref. \citenum{Lee10} remains unchanged in the final LRS2 design, there have been some major adjustments. The largest is that the quad-channel simultaneous coverage has been broken up into two separate double-spectrographs that will independently observe the wavelength ranges of $370 < \lambda (\mathrm{nm}) < 700$ and $650 < \lambda (\mathrm{nm}) < 1050$, respectively. The blue optimized spectrograph pair (LRS2-B) requires only modest adaptation of the VIRUS design while the red optimized pair (LRS2-R) essentially requires the same modifications as LRS2-B in addition to significant work to adapt the VIRUS camera to the differently packaged red sensitive, thick deep-depletion CCDs. When complete, the latter will significantly improve the performance as compared to the current LRS. A significant factor in dividing LRS2 into two separate double-spectrographs was to ensure the continuity of capable low resolution spectroscopy on the HET through the WFU. As such, LRS2-B will be ready for commissioning during quarter 4 of 2013 so that it is available immediately for the upgraded HET. 

As stable instruments with rapid setup times and high efficiency, LRS2-B and LRS2-R are designed to be highly complementary to VIRUS and to exploit the HET queue-scheduling for survey follow-up, synoptic observations, and targets of opportunity. Since LRS2-B is planned to be a commissioning instrument for the HET WFU, we focus in this work on presenting the details of its final design. In $\S$\ref{sec:science}, we describe the science motivation for LRS2 with an emphasis on the drivers for setting the spectral resolution and wavelength coverage of each of the two LRS2-B channels. In $\S$\ref{sec:design}, we outline the design of the instrument and focus on the departures from the nominal VIRUS design. In $\S$\ref{sec:status}, we discuss the current status of the instrument, including the progress in procuring parts, its construction, and lab testing. In $\S$\ref{sec:lrs2-r}, we change our focus to LRS2-R, and give a brief review of its science drivers as well as the additional details of its design that depart from that of LRS2-B. Finally, we briefly discuss the plan for completing the instrument in $\S$\ref{sec:outlook}.

\section{SCIENCE MOTIVATION}\label{sec:science}
LRS2 will play to the strengths of the HET and be an excellent replacement to continue the current science usage of LRS. While a VIRUS-based design will not replicate the multi-object or imaging capabilities of the current LRS, the majority of observations have been spectroscopic for single small objects ($<10$\arcsec\ in size). Additionally, the imaging capability will be assumed after the WFU by the new acquisition camera\cite{Vattiat12} and the multi-object spectroscopy will be covered by VIRUS itself\cite{Hill12a}. The strongest driver for LRS2 is efficient broadband coverage with high efficiency, particularly in the red. The final LRS2 instrument with separate complimenting blue and red optimized instruments will deliver high sensitivity and excellent sky subtraction through the new IFU. LRS has historically been the dominant instrument in dark time and has had the largest variety of science applications of all three current HET instruments. We expect LRS2 to have similar wide applications and to increase the efficiency of some of the tasks at which its versatile predecessor excelled. In this section, we outline some of the science motivation that has driven the design of LRS2-B. Many additional intriguing science topics can also be explored with the enhanced red sensitivity and increased bandwidth when LRS2-B and LRS2-R are available together. We discuss such science topics later in $\S$\ref{subsec:lrs2-rscience}.

\subsection{LRS2-B as a Facility Instrument}\label{subsec:lrs2-bgeneral}
As a replacement for LRS and as a second generation facility instrument, LRS2 must enable all of the ongoing LRS science and provide new science opportunities for the HET community. Such general use capabilities are a major driver for the instrument design. LRS2-B, assuming the red end of its spectral coverage ends at $\sim700$ nm, will extend and improve upon the usage of LRS in many science topics. As mentioned previously, the versatility of LRS coupled with the queue scheduling of the telescope have led it to being an ideal instrument for targets of opportunity and for time critical studies of transient phenomena, such as follow-up spectroscopy of $\gamma$-ray bursts\cite{Schaefer03} and supernovae (both for single object studies of supernovae physics\cite{Quimby07} and in survey mode for classification\cite{Zheng08}). Similar future observations will greatly benefit from the LRS2 IFU since it will significantly decrease the setup time and increase observing efficiency since coordinates of transients often arrive with arcsecond-level errors. The IFU will also improve the 2-D background subtraction of the underlying host galaxies of $\gamma$-ray bursts and supernovae. Additionally, synoptic observations, such as AGN reverberation mapping\cite{Kaspi07}, have been well suited to the HET/LRS capability and will continue to be with LRS2-B. LRS has also contributed to studies of galaxy structure and dynamics\cite{Corsini08} as well as identification and studies of high-$z$ quasars\cite{Schneider00} ($z \lesssim 4$ for the assumed LRS2-B wavelength coverage). LRS2-B will additionally continue to enable emission line diagnostics in the local universe for the standard optical transitions (e.g., \oii, \oiii, \hb, \ha, \nii) and will have the added advantage over LRS of providing a modest amount of 2-D spatial information that subsamples the HET median image quality ($\sim1.2$\arcsec\ FWHM).

With wide wavelength range, excellent sky subtraction, and efficient operation in survey mode when coupled with the queue scheduled HET, LRS2 will play a uniquely important role as a follow-up instrument for interesting objects in large surveys. LRS has already proven to be a useful instrument in the identification and study of optical counterparts to objects discovered in multi-wavelength imaging surveys (e.g., radio\cite{Brand03} and X-ray\cite{Hornschemeier03} surveys). Similarly, LRS2 will prove to be complimentary to upcoming major imaging surveys, such as the Dark Energy Survey, Pan-STARRS, and LSST, that will discover an unprecedented number of objects requiring spectroscopic follow-up. Additionally, LRS2 will be highly complementary to the HETDEX survey in which VIRUS will discover almost a million LAE at $1.9 < z < 3.5$, a similar number of \oii\ emitting galaxies for $z<0.5$ in addition to interesting stars, clusters of galaxies, high-$z$ quasars, AGN, and more down to the sensitivity limit of the SDSS imaging survey. This unique dataset will require follow-up with LRS2 to establish the classification of emission line objects (e.g., LAE, \oii\ emitters, and AGN), to classify extremely metal poor stars, and to study emission line diagnostics over a wider wavelength range (since VIRUS only can observe to 550 nm). Such an example of the latter capability is the study of galactic winds, which is ripe for advancement over the next decade. HETDEX will provide a massive sample of blue compact dwarf galaxies at $z<0.5$ where LRS2-B can search for the emission-line diagnostics of outflows. Finally, given the large sample resulting from HETDEX, LRS2-B will be the ideal instrument for following-up the most interesting high-$z$ star-forming galaxies, particularly LAE. Since this science is a major driver of the instrumental design, we outline it in more detail in the following section. 

Note that while LRS2-B will be a powerful instrument in its own right for carrying out the observations featured in the non-exhaustive list of science topics discussed here, LRS2-R will extend all of its capabilities to higher wavelengths. This  will allow extragalactic studies to reach higher $z$ and enable additional science with the enhanced red sensitivity over the current LRS. This is further discussed in $\S$\ref{subsec:lrs2-rscience}

\subsection{Star-Forming Galaxy Physics at $z\sim2.5$}\label{subsec:lrs2-blya}
LRS2-B will be the ideal instrument to follow-up the most interesting LAE discovered by HETDEX (e.g., largest mass, highest star formation rate, etc.) at the epoch when the star formation rate of the universe was near its peak and when most galaxy assembly occurred. Obtaining higher resolution spectra over large wavelength coverage will allow \lya\ line profiles and other emission lines (e.g. CIV) to be studied, probing the interstellar medium of these galaxies. The transfer of \lya\ photons has long been an intense field of study largely because \lya\ emission is a primary selection technique of high-$z$, star-forming galaxies. However, \lya\ radiative transfer is complicated due to the resonant nature of the $n=2\rightarrow1$ electronic transition of hydrogen, which effectively scatters \lya\ photons through neutral hydrogen in a geometric and frequency diffusion process that terminates upon the photon reaching the line wing where the optical depth is sufficiently low for escape. The radiative transfer is extremely sensitive to the complex velocity fields of the gas as well to the presence and distribution of dust due to the photon's long effective path length through the galaxy. This information is thus encoded in the observed \lya\ line profile, and results in a spread of complex \lya\ emission line morphologies which range from single asymmetric peaks to multiple peaks with varying asymmetries distributed around the intrinsic line center. Various state-of-the-art simulations\cite{Verhamme06,Barnes11} have been published and attempt to explain \lya\ radiative transfer in the context of star-forming galaxies. New observations\cite{Kulas12} are beginning to result in data sets that will be useful for comparison to and placing constraints on such models. This could be extremely useful in learning about the early evolution of galaxies, determining the gas velocity fields of star-forming galaxies, as well as learning about the effects of feedback (e.g., star burst driven winds) from extreme star formation on the galactic environment.

   \begin{figure}[b]
   \begin{center}
   \begin{tabular}{c}
   \includegraphics[width=0.98\textwidth]{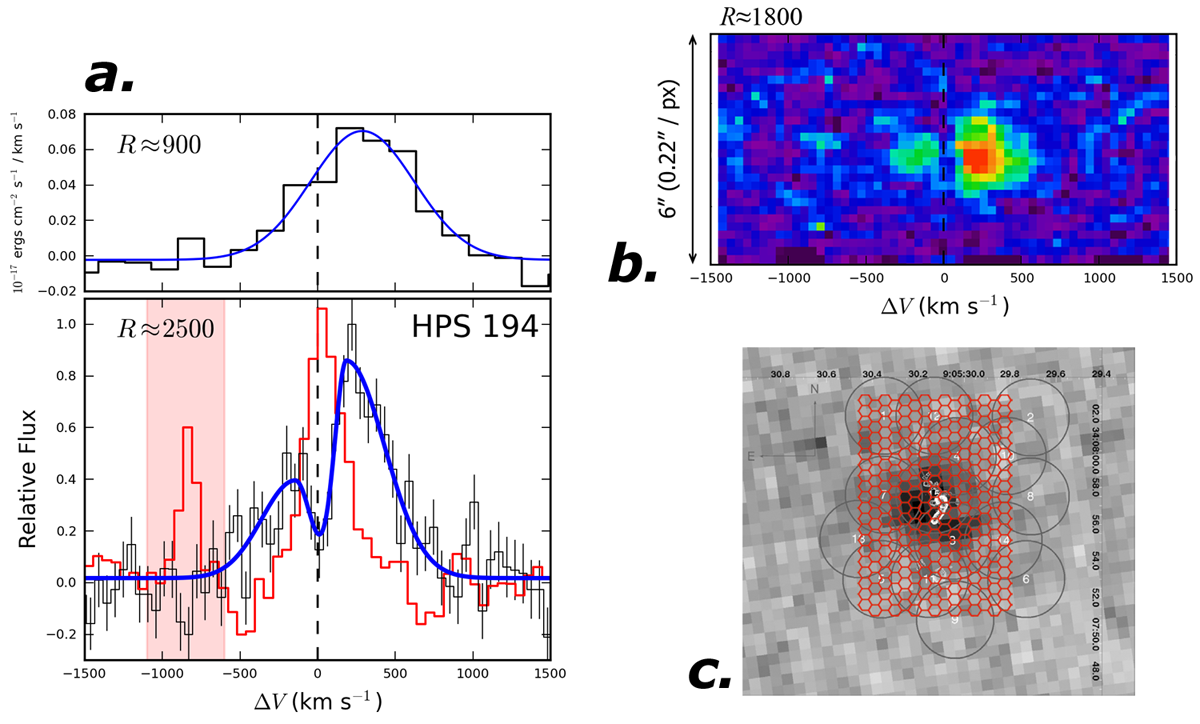}
   \end{tabular}
   \end{center}
   \caption[example] 
   { \label{fig:LAEScience} 
\textit{a}) \lya\ spectra of HPS 194, a $z \approx 2.3$ LAE discovered in the HETDEX Pilot Survey with the Mitchell Spectrograph (i.e., the VIRUS prototype spectrograph) at the McDonald Observatory 2.7 m telescope from Ref. \citenum{Chonis12}. The top panel shows the low resolution discovery spectrum\cite{Adams11} and the lower panel shows the same galaxy observed with a higher resolution grating in which multiple emission components are observed. The red spectrum is a NIR observation of \ha\ from Ref. \citenum{Finkelstein11} that has been arbitrarily scaled to the peak \lya\ flux. The \lya\ spectra have been shifted to the centroid of the optically thin \ha\ line. The blue curves are parametric fits to the \lya\ data. \textit{b}) A 2-D slit spectrum of HPS 194 taken with the IMACS instrument from Ref. \citenum{Chonis12}, which shows the multiple \lya\ peaks that are spatially resolved (the seeing was 0.8\arcsec\ FWHM). The spectral resolution is similar to that of the UV Arm of LRS2-B. \textit{c}) A \lya\ image of the $z = 3.4$ radio galaxy B2 0902+34 from Ref. \citenum{Adams09}, which shows an extended \lya\ halo. The gray circles represent the 4.2\arcsec\ diameter fibers of the Mitchell Spectrograph IFU in which \lya\ was detected. The red overlay shows the LRS2-B IFU (see $\S$\ref{subsec:IFUfeed}), which can perform much finer spatial sampling for detailed studies of extended \lya\ emission. 
}
   \end{figure} 

LRS2-B will be able to significantly contribute to this field of study and this science has driven the design of the instrument. The redshift range $2.0 \lesssim z \lesssim 2.7$ ($365 \lesssim \lambda (\mathrm{nm}) \lesssim 450$ for \lya) is of particular interest because the galaxies' rest-frame optical nebular transitions (e.g., \hb, \oiii, \ha, \nii) redshift into the near-infrared (NIR) atmospheric windows. Since these transitions are optically thin, observing\cite{McLinden11,Finkelstein11} them in conjunction with the \lya\ line can provide a measure of the systemic \lya\ line center and intrinsic line emitting gas velocity dispersion, which are critical for constraining\cite{Kulas12} the \lya\ radiative transfer models. Spectral resolution is important in order to resolve the velocity structure in the \lya\ emission line profile. For continuum selected galaxies that show \lya\ in emission, the separation between multiple emission components (when observed) is typically on the order 800 km/s\cite{Kulas12}. However, the velocity offsets from the systemic line center that are observed for LAE\cite{McLinden11} are often a factor of $\sim2$ smaller than for continuum selected galaxies, and recent observations\cite{Chonis11,Chonis12} have shown that this is also true for the separation between multiple emission components (when observed). Thus, being able to resolve emission components separated by $\lesssim300$ km/s is essential for this work, as it will help to identify and interpret line profile morphologies. To highlight this, Fig. \ref{fig:LAEScience} shows the \lya\ spectra from Ref. \citenum{Chonis12} of a bright LAE discovered in the HETDEX Pilot Survey (HPS\cite{Adams11}) observed at different spectral resolutions with the Mitchell Spectrograph (formerly VIRUS-P\cite{Hill08b}) on the McDonald Observatory 2.7 m Harlan J. Smith telescope ($a$) and the IMACS\cite{Dressler11} instrument on the 6.5 m Magellan Baade telescope ($b$). Utilizing the large sample of LAE from HETDEX, LRS2-B will efficiently be able to obtain higher resolution spectra for a large number of interesting LAE discovered by VIRUS. With a new generation of NIR multi-object spectrographs soon coming on-line (e.g., KMOS\cite{Sharples12} and MOSFIRE\cite{McLean12}), a full suite of data will become available for interpreting \lya\ line profiles with statistically significant sample sizes. While similar work has recently begun for continuum selected galaxies\cite{Kulas12}, doing so for LAE will also be especially important since LAE are by definition the highest \lya\ equivalent width sources and are intrinsically a different sample of galaxies. 

Among the most interesting LAE that HETDEX will discover and require subsequent LRS2-B follow-up of are the spatially extended LAE, or \lya\ blobs. These objects, with sizes as large as 100 kpc, can be studied in great detail with the LRS2-B IFU.\footnote{At $z = 2.5$, the 7\arcsec$\times$12\arcsec\ LRS2-B IFU subtends $57\times98$ kpc assuming a standard cosmology.} In the HPS sample\cite{Adams11}, five out of the 105 emission line sources that were identified as \lya\ were spatially extended (i.e., had sizes of $>7$\arcsec, which is the spatial resolution limit of the survey). With the higher spatial sampling of VIRUS on the HET for HETDEX, this detection rate suggests that HETDEX will identify extended \lya\ sources in unprecedented numbers, thus providing a large sample from which to choose follow-up targets. Such objects may be sites of intense in-fall of gas\cite{Adams09} or strong wind outflows and are important zones of interaction between galaxies (or groups of galaxies\cite{Yang10}) and the surrounding intergalactic medium. Such studies will benefit greatly from the 3-D spectroscopy LRS2-B can provide through the analysis of the \lya\ line profile velocity structure as a function of position on the sky, which may help determine the emission mechanism of these enormous objects (e.g., star formation, radiative feedback from AGN, gravitational cooling, or some combination). As shown in see Fig. \ref{fig:LAEScience}$c$, the IFU size and sampling will be well suited to such studies even for the lower redshifts for which \lya\ will be accessible with LRS2-B. In addition to the \lya\ blobs, LRS2-B will be useful for studying the extended nature of the \lya\ emission around single LAE, as evidenced by the 2-D slit spectrum in Fig. \ref{fig:LAEScience}$b$.

One of the advantages LRS2-B has as a standalone instrument as compared to VIRUS is that its wavelength coverage extends $\sim150$ nm redward to $\sim700$ nm. This allows for the potential to study additional features of star-forming galaxies in the rest-frame ultraviolet (UV) that lie significantly redward of \lya\ (such as CIV). Additionally, this extends the redshift range for which \lya\ can be observed to $z \approx 4.75$. With the queue scheduled HET and through its parallel observing mode\cite{Odewahn12} that will be implemented through the WFU (where the ability to simultaneously observe with VIRUS and another HET instrument is enabled), one could efficiently carry out a survey for LAE in congruence with HETDEX to higher redshift that would be similar in size to the HPS for extending studies of the evolution of LAE properties\cite{Blanc11} to earlier epochs. 

\section{LRS2-B INSTRUMENT DESIGN}\label{sec:design}
The design of LRS2 is based on the VIRUS unit spectrograph and benefits greatly from the large investment made in its design, prototyping, and production line engineering. We begin by discussing the key design features of VIRUS (a full description of the instrument can be found in Ref. \citenum{Hill12a}) and continue by outlining the design of the modifications and adaptations that morph VIRUS into LRS2-B.

\subsection{VIRUS: The Building Block of LRS2}\label{subsec:VIRUS}
   \begin{figure}[t]
   \begin{center}
   \begin{tabular}{c}
   \includegraphics[width=0.98\textwidth]{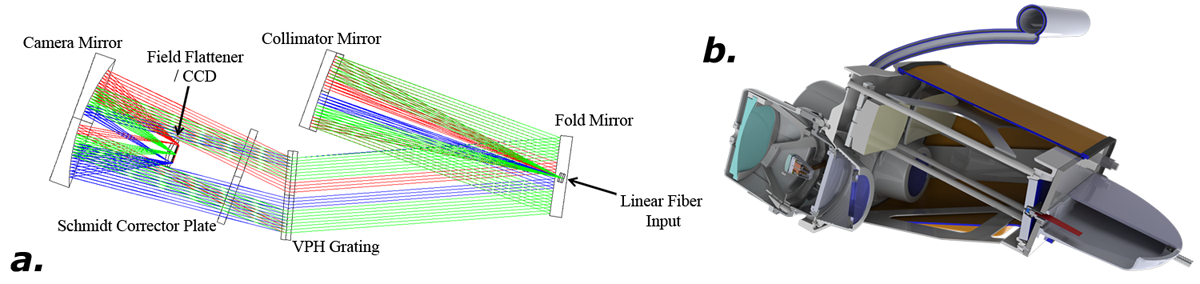}
   \end{tabular}
   \end{center}
   \caption[example] 
   { \label{fig:VIRUS} 
\textit{a}) The optical layout of VIRUS with major components labeled. This layout is the same as for the LRS2-B channels except that the VPH gratings are exchanged for a grisms (see Fig. \ref{fig:GrismSchem}) and the field flattener design in the Orange Arm is slightly different. For scale, the collimated beam size is 125 mm. The fiber ``slit'' is oriented out of the page. \textit{b}) A section view of the mechanical design of the VIRUS unit spectrograph pair. The plumbing extending above the spectrograph is part of the liquid nitrogen distribution manifold\cite{Chonis10}. The fiber ``slit'' is mounted to the bulkhead at right. 
}
   \end{figure} 
The VIRUS unit spectrograph is designed to be fed by fiber optics and is optimized to accept a $f$/3.65 input beam (though it can accept a beam as fast as $f$/3.4 to account for a small amount of focal ratio degradation in the fibers\cite{Murphy08} and misalignments). The optical design (see Fig. \ref{fig:VIRUS}$a$) is a double-Schmidt (where the collimator is reversed) with a shared corrector plate and utilizes a volume phase holographic (VPH) grating for dispersion at the pupil. The focal reduction factor is 2.8. The primary wavelength coverage for HETDEX is the near-UV to green region of the spectrum, but the design is pan-chromatic since the optical power is mainly in the two spherical mirrors. For adaptation to different wavelength ranges while maintaining the superb image quality of the nominal design, the mirrors and transmissive optics can be supplied with different coatings and custom corrective optics can be fabricated (see $\S$\ref{subsec:camera} and \ref{subsec:lrs2-rdesign}). The dispersion and wavelength coverage can be further adapted by using grisms instead of a standard VPH grating (see $\S$\ref{subsec:grisms}). VIRUS is designed to be fed by a "densepak" IFU, where the fibers are arrayed in a hexagonal pattern with 1/3 fill-factor at the input from the telescope. For input into the instrument, the fibers are arranged in a linear ``slit'' (see $\S$\ref{subsec:IFUfeed}). The VIRUS IFU is fed directly at $f$/3.65 by the HET Wide Field Corrector (WFC). An observation with VIRUS consists of three exposures with small offsets (i.e., dithers of $\sim1$\arcsec) that fill in the sky coverage gaps between the fibers. The dithers are achieved by precisely moving the focal plane assembly on which the IFUs are mounted rather than moving the telescope itself (see Ref. \citenum{Vattiat12}). The dithering is thus independent of guiding and guarantees precise knowledge of the relative positions of the three exposures that complete an observation. In production, each VIRUS mechanical unit contains a pair of spectrographs in a common housing (see Fig. \ref{fig:VIRUS}$b$). Since this configuration allows a pair of spectrographs to share common components (such as CCD electronics, vacuum vessels, cryogenic components, etc.), costs are reduced and more efficient use of the limited space available at the telescope structure is allowed for. Two of these unit pairs will form the basis of LRS2: one for LRS2-B and one for LRS2-R. We have constructed a prototype single VIRUS spectrograph (the Mitchell Spectrograph\cite{Hill08b}), which has been used on the McDonald Observatory 2.7 m telescope since 2007 to verify the performance of the instrument and to complete the HPS\cite{Adams11} in support of HETDEX. The Mitchell Spectrograph has demonstrated high throughput, excellent image quality, very low ghosting, and exceptional stability. The result has been evolved into the final replicated units for production. 

\subsection{Feed Optics and Integral Field Units}\label{subsec:IFUfeed}
   \begin{figure}[t]
   \begin{center}
   \begin{tabular}{c}
   \includegraphics[width=0.8\textwidth]{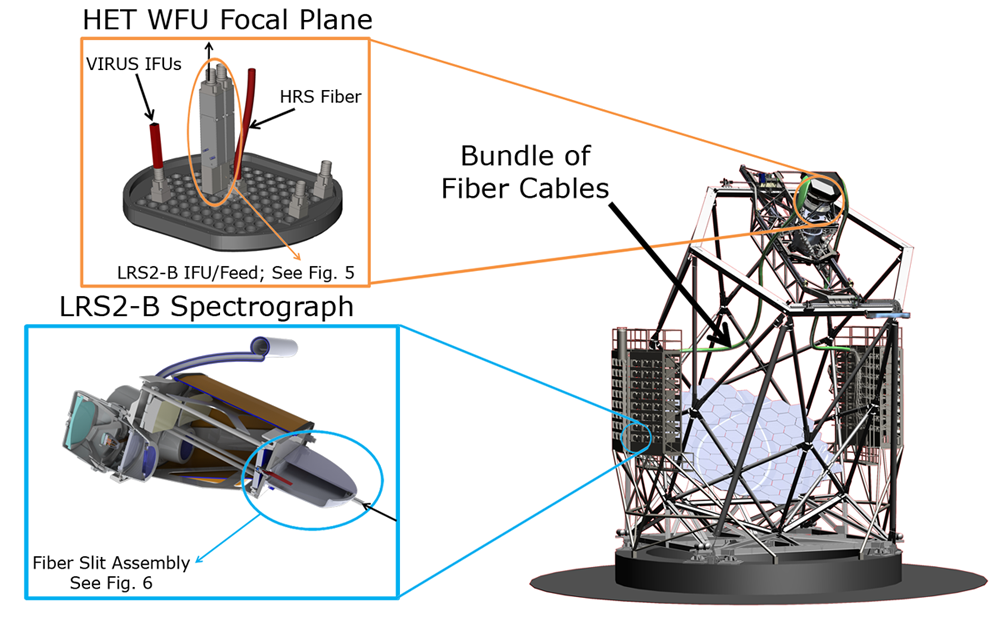}
   \end{tabular}
   \end{center}
   \caption[example] 
   { \label{fig:FiberRoute} 
An outline of the front and back ends of the LRS2-B fiber light path. The HET focal plane where the VIRUS IFUs are arrayed is located at the top of the telescope inside the Prime Focus Instrument Package enclosure. The focal plane is schematically detailed in the orange inset, in which the LRS2-B IFU feed enclosure can be seen mounted in the center of the focal plane with a few VIRUS IFUs installed around it. This highlights the limited space available in which the LRS2-B feed optics must be deployed. The IFU and feed optics are detailed in Fig. \ref{fig:IFUFeed}. At the other end of the fiber cable is the spectrograph pair shown in the blue inset, where the fibers that were arrayed at the input IFU end are laid out in a linear slit for input into the spectrograph (see Fig. \ref{fig:FiberSlit}). Arrowed in the rendering of the HET at right is the large bundle of fiber cables in which the fibers from each IFU are directed towards the support structures in which VIRUS and LRS2 are housed. 
}
   \end{figure} 
The largest departure from the VIRUS design for LRS2-B is the IFUs and how they are fed from the telescope. The bare 266 $\mu$m core diameter fibers of the VIRUS IFU (which have an on-sky diameter of 1.5\arcsec) are fed directly by the HET WFC. For LRS2-B, we desire an IFU with close to unity fill-factor that subsamples the typical HET image quality, which will require reimaging the telescope's incoming beam. Fig. \ref{fig:FiberRoute} outlines the front and back ends of the LRS2-B instrument's fiber light path (i.e., the feed optics/IFU and the fiber slit on the spectrograph pair, respectively).

\begin{figure}[t]
   \begin{center}
   \begin{tabular}{c}
   \includegraphics[width=0.98\textwidth]{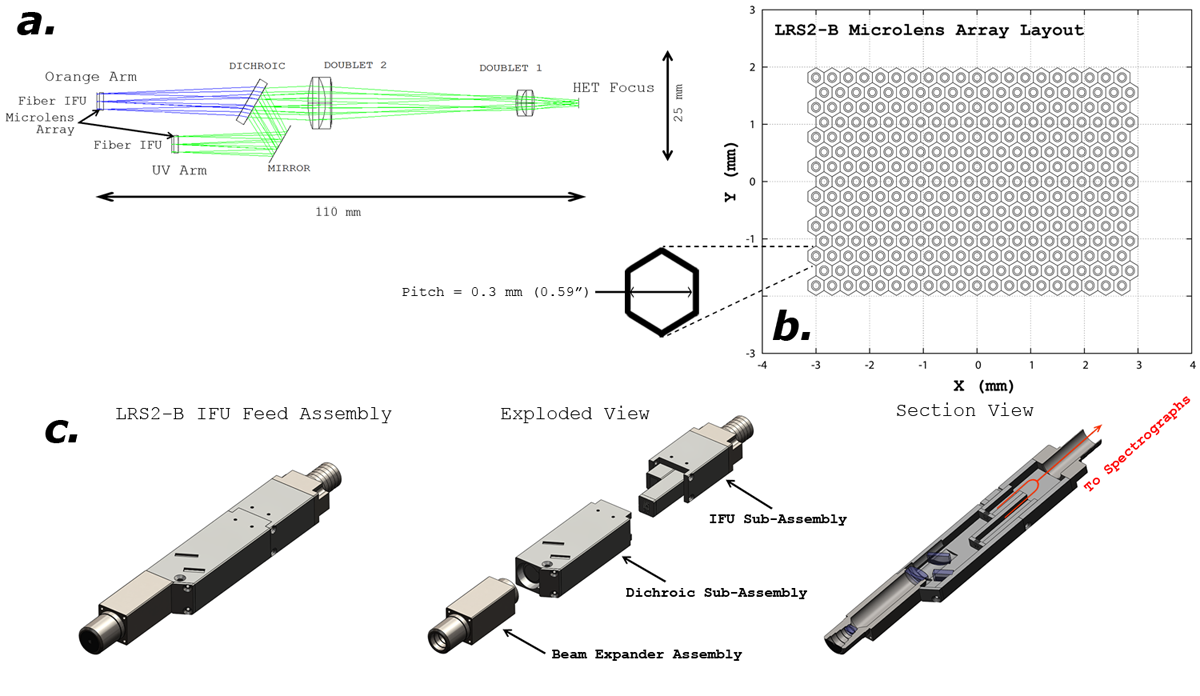}
   \end{tabular}
   \end{center}
   \caption[example] 
   { \label{fig:IFUFeed} 
The LRS2-B IFU feed design. \textit{a}) Optical design of the IFU feed. The HET focus is at right and the diverging beam is collimated by Doublet 1 and subsequently reimaged at $f$/10.5 by Doublet 2. The light is split between the UV and Orange Channels by the custom dichroic. For each channel, the beam is focused on a microlens array, which forms micro-pupil images on the IFU fiber ends. \textit{b}) Layout of the two LRS2-B $15\times20$ hexagonal element microlens arrays. The larger circle schematically shown at the center of each microlens represents a fiber core of the IFU while the smaller circle represents the micro-pupil size. The zoomed view of a single hexagonal spaxel defines the pitch size. \textit{c}) Three views of the IFU feed mechanical design, which is built into three main sub-assemblies. On the right, the red arrows indicate the path of the fiber bundles from each IFU, which combine into a single bundle before being routed into conduit. The fibers are not explicitly shown. The fibers for each channel are divided to each respective spectrograph as shown in Fig. \ref{fig:FiberSlit}.
}
   \end{figure}

The LRS2-B IFU feed optics (see Fig. \ref{fig:IFUFeed}) consist of a focal expander made with two custom doublets (consisting of fused silica and LLF1 glass) mounted after the telescope focus that convert the $f$/3.65 telescope beam to $f$/10.5 for finer spatial sampling. The field of view of the focal expander is $\sim14$\arcsec\ in diameter. An optimized dichroic after the reimaging doublet (i.e., Doublet 2 in Fig. \ref{fig:IFUFeed}$a$) reflects blue light ($\lambda < 466$ nm), while transmitting red light ($\lambda > 466$ nm). The fold mirror in the blue arm will have a UV/blue enhanced coating. The dichroic is used at 30\degree\ angle of incidence to help sharpen the the transition between the transmission and reflection regions. Note that the details of the choice of transition wavelength for the dichroic are discussed in the following section. The image quality of the dual-doublet relay is excellent with the diameter enclosing 90\% of the diffraction encircled energy being $<0.25$\arcsec\ across the central 12\arcsec\ of its field of view. 

\begin{table}[b!]
\caption{LRS2-B Feed Optics and IFU Properties} 
\label{tab:IFUProp}
\begin{center}       
\begin{tabular}{|c|c|c|c|c|c|c|} 
\hline
\rule[-1ex]{0pt}{3.0ex} \footnotesize{\textbf{Dichroic}} & \footnotesize{\textbf{Feed Focal}} & \footnotesize{\textbf{Lenslet}} & \footnotesize{\textbf{Micro-Pupil}} & \footnotesize{\textbf{Fiber Core}} & \footnotesize{\textbf{Lenslet}} & \footnotesize{\textbf{Spaxel Center-to-}}  \\

\rule[-1ex]{0pt}{3.0ex} \footnotesize{\textbf{Crossover}} & \footnotesize{\textbf{Ratio}} & \footnotesize{\textbf{Pitch}} & \footnotesize{\textbf{Diameter}} & \footnotesize{\textbf{Diameter}} & \footnotesize{\textbf{Array Format}} & \footnotesize{\textbf{Center IFU Size}} \\

\rule[-1ex]{0pt}{3.5ex} \footnotesize{(nm)} & & \footnotesize{(mm)} & \footnotesize{($\mu$m)} & \footnotesize{($\mu$m)} & & (\arcsec) \\
\hline
\rule[-1ex]{0pt}{0.8ex}  & & & & & & \\
\rule[-1ex]{0pt}{1.8ex}  \footnotesize{$466.0\pm1.5$} & \footnotesize{$f$/10.5} & \footnotesize{0.30 (0.59\arcsec)} & \footnotesize{120} & \footnotesize{170} & \footnotesize{$15\times20$} & \footnotesize{$7.1\times11.5$} \\
\rule[-1ex]{0pt}{1.8ex}  & & \footnotesize{Hexagonal} & & & \footnotesize{Rectangular} & \\
\hline
\end{tabular}
\end{center}
\end{table}

\begin{figure}[t]
   \begin{center}
   \begin{tabular}{c}
   \includegraphics[width=0.85\textwidth]{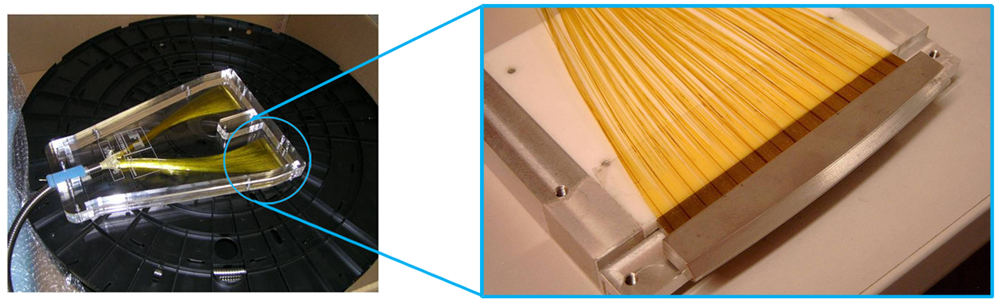}
   \end{tabular}
   \end{center}
   \caption[example] 
   { \label{fig:FiberSlit} 
The fiber ``slit'' at the spectrograph end of the fiber cable. The fibers from each IFU in the feed assembly are routed to the spectrographs through the same fiber cable which terminates at the assembly shown in the left panel. The fibers are separated into a double slit arrangement before integration into a housing that mounts onto the spectrograph pair's mechanical framework. The zoomed in view shows the slit end for one of the spectrographs in a production jig. This slit end of the LRS2-B fiber cabling is identical to that used for VIRUS.
}
   \end{figure}

LRS2-B will have a near-unity fill factor IFU as driven by the science outlined in $\S$\ref{sec:science}. Such an IFU will help to ease target acquisition and guarantee adequate sky coverage in all image quality conditions. This is achieved by coupling the focal plane of the focal expander to the IFU fibers through a microlens array, which forms micro-pupil images on the fiber ends and can provide a spatial fill-factor well above 95\%\cite{AllingtonSmith02}. For LRS2-B, we place a microlens array at each of the two foci of the focal expander, which then feed an IFU for each of the two spectrograph channels (see Fig. \ref{fig:IFUFeed}$a$). To help improve the spectral resolution over VIRUS, the IFUs will be constructed with fibers that have 36\% smaller core diameter (i.e., 170 $\mu$m). The microlens array has hexagonal elements with a pitch of 0.3 mm (see Fig. \ref{fig:IFUFeed}$b$). The pitch size corresponds to 0.59\arcsec\ on the sky, which allows optimal sampling of the nominal seeing disk at HET (1.2\arcsec\ FWHM). The microlens/IFU layout is $15\times20$, which covers a 7.1\arcsec$\times$11.5\arcsec\ field as measured from the centers of the microlens elements. The elongated field is advantageous since it places potential sky fibers farther from an observed object as compared to a circular microlens arrangement. The microlens array converts the $f$/10.5 expanded beam back to $f$/3.65 onto the IFU fiber cores for optimally feeding the spectrograph. The micro-pupil size produced by each microlens is 0.12 mm in diameter, which gives 0.05 mm of leeway for the IFU-to-microlens array alignment to minimize light loss. The microlens coupling to the fibers is particularly effective for the HET since the azimuthal scrambling by the fibers provides for a stable output beam into the spectrograph that is independent of the illumination of the telescope pupil (which changes as the HET tracks). However, light losses are slightly higher in this situation as compared to direct imaging onto the fibers. This is because today's high quality optical fibers exhibit little enough focal ratio degradation when fed at $\sim f$/3.6 that the telescope's central obstruction is maintained in the spectrograph for directly illuminated fibers (such as for VIRUS). Since obscurations exist in the spectrograph due to its folded collimator and Schmidt camera, the loss of the telescope central obstruction in the pupil that is formed within the sprectrograph results in extra losses when feeding the fibers with a micro-pupil by way of the microlens. Once each IFU is fed, the fibers for each IFU are bundled together through a single conduit and routed to the spectrograph pair. As shown in Fig. \ref{fig:FiberSlit}, the fibers are then again separated and arranged into a linear ``slit'' for input into each respective spectrograph channel. In Table \ref{tab:IFUProp}, we summarize the properties of the fiber feed optics and IFU.

As shown in Fig. \ref{fig:FiberRoute}, the LRS2-B IFU feed is integrated directly into the center of the same focal plane assembly on which all of the VIRUS IFUs are mounted and the other HET facility instruments (e.g., HRS and MRS) also reside. This allows the implementation of the highly desired parallel observing mode\cite{Odewahn12} where VIRUS can observe in the background while another HET instrument is observing (see more regarding observing with LRS2-B in parallel mode in $\S$\ref{subsec:operation}). While advantageous, mounting the LRS2-B IFU feeds on the same focal plane assembly as VIRUS presents a design challenge, largely due to the limited space available for the feed optics. The preliminary mechanical design of the IFU feed is shown in Fig. \ref{fig:IFUFeed}$c$. It consists of three sub-assemblies: the first contains the two beam expander doublets, the second contains the dichroic and fold mirror, and the third contains the microlens arrays and IFUs. Based on our optical tolerance analysis, most of the optics can be mounted in the mechanical assembly to machine accuracy. However, we are currently in the process of engineering the compensator adjustment into the design, which is challenging due to the space constraints. The driving tolerance is two-fold for the successful fabrication and alignment of the IFU feed optics. First, 1) the registration of the two IFUs must be accurate for spatially resolved studies of extended objects for which the broad wavelength coverage of the combination of the two spectrograph channels is desired. The relative image shift should be $< 1/10$ of the 0.3 mm lenslet pitch. Additionally, 2) the proper focal ratio at the input of the IFU fibers needs to be maintained slower than $f$/3.4, which is the limit that the spectrograph can operate without the input beam overfilling the spectrograph pupil. We are in the process of developing the final fabrication and alignment error budget for the optical and mechanical components.

   \begin{figure}[t]
   \begin{center}
   \begin{tabular}{c}
   \includegraphics[width=0.6\textwidth]{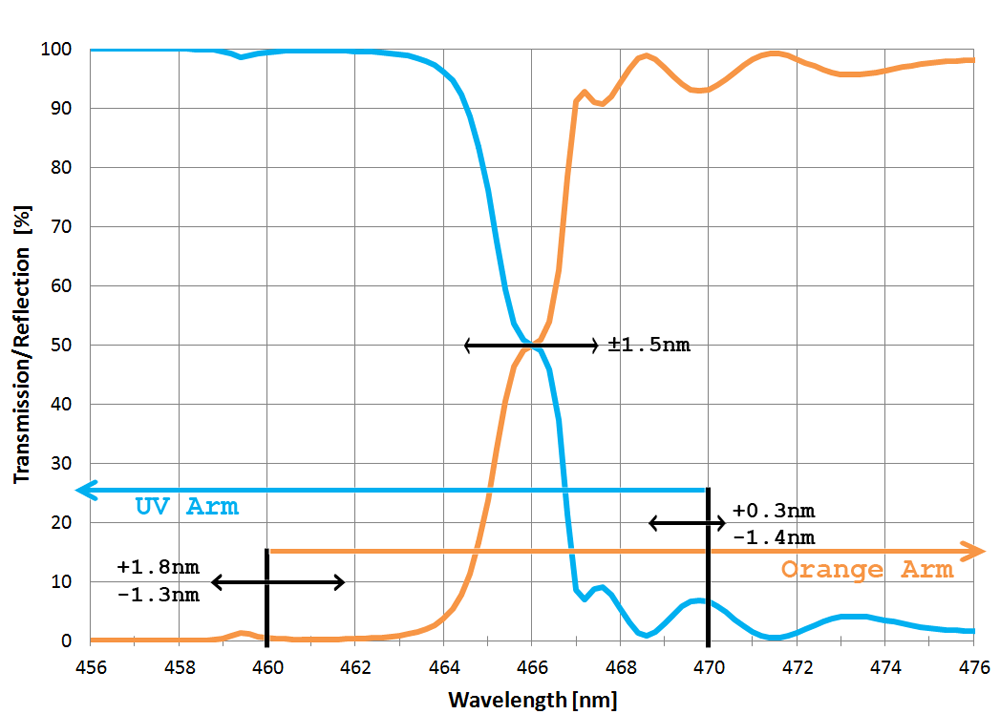}
   \end{tabular}
   \end{center}
   \caption[example] 
   { \label{fig:DichroicCross} 
A plot of the dichroic reflectance (for the UV Arm) and transmittance (for the Orange Arm) in the transition region between the two LRS2-B channels as provided in the design proposal by JDSU Corp. Also shown are the ranges of variation expected in the dichroic's transition wavelength and the edges of each channel's wavelength coverage that results from the tolerance on each VPH grism (the latter is determined from a Monte Carlo analysis). 
}
   \end{figure} 

\subsection{VPH Grisms}\label{subsec:grisms}
Since LRS2-B is housed inside the mechanical framework of a VIRUS spectrograph, the collimator and camera angles are fixed. To adapt such a setup to a wavelength coverage configuration and spectral resolution that differs from the nominal VIRUS design for a fixed fiber core diameter, one can simply immerse a standard VPH grating between prisms (i.e., a grism) to tune the beam deviation for various grating fringe frequencies. Note that the fixed design naturally creates a trade-off between the spectral coverage and resolution, thus the design of the VPH grisms is strongly dependent on the desired wavelength coverage for each spectrograph channel. As discussed in $\S$\ref{sec:science}, emission line diagnostics in the local universe has historically been a significant use of LRS. To maintain this, the transition region between the two spectrograph channels should avoid the spectral region around \heii, \hb\, and \oiii\ for low redshift while accommodating the broad line widths often observed in AGN. Additionally to enable LAE follow-up, the redshift range in which the important rest-frame optical transitions are shifted into the NIR atmospheric windows should be accommodated. These considerations place the channel transition between 460 and 470 nm. An additional important emission line to consider is \oii\ in the local universe; including this transition sets the lower wavelength bound to be 370 nm, which extends the capability as compared to the current LRS further toward the UV by $\sim50$ nm. We thus call the bluer LRS2-B channel the ``UV Arm''. A commonly used mode of LRS is the ``G2'' configuration, which simultaneously observes from below \hb\ to \ha. To replicate this mode and account for some moderate range in redshift, we set the upper wavelength bound to 700 nm. The redder LRS2-B channel is thus called the ``Orange Arm''. To maximize the useful wavelength coverage of each channel, the dichroic transition must be as sharp as possible. Dialog with vendors have shown that a 2 nm transition from 20\% to 80\% transmittance/reflectance is possible for a 50\% transition wavelength of $466.0\pm1.5$ nm (see Fig. \ref{fig:DichroicCross}).

   \begin{figure}[t]
   \begin{center}
   \begin{tabular}{c}
   \includegraphics[width=0.8\textwidth]{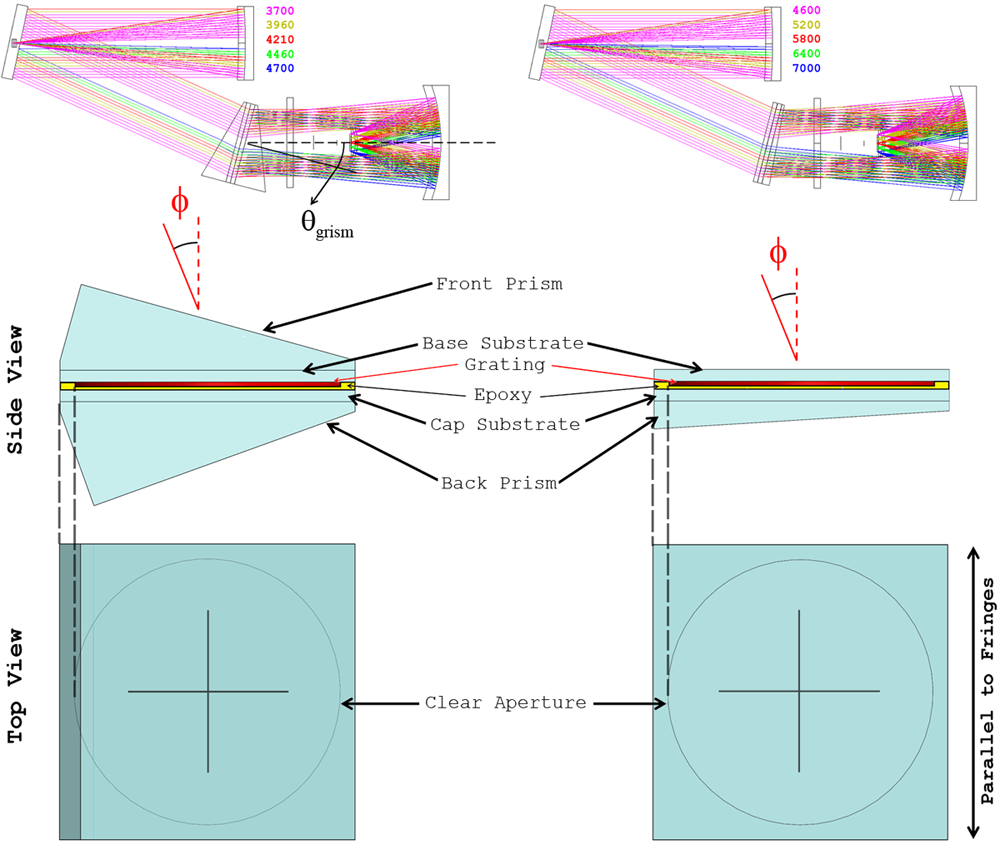}
   \end{tabular}
   \end{center}
   \caption[example] 
   { \label{fig:GrismSchem} 
The LRS2-B grisms for the UV Arm (left column) and Orange Arm (right column). The top row shows a ray trace of each spectrograph channel. Superposed on the UV Arm schematic is a dashed line corresponding to the camera's optical axis. The solid line is the normal to the diffracting layer. The angle between these, $\theta_{\mathrm{grism}}$, is what we call the grism assembly physical tilt. The middle row shows a side view of each grism with major components labeled. In this view, the fringes are oriented into the page and the direction of the fringe tilt is labeled as $\phi$. The grating layer thickness is not shown to scale. The bottom row shows a top view of each grism, highlighting the square physical footprint to ease mounting and the circular (140 mm diameter) clear aperture.  
}
   \end{figure} 
\begin{table}[b!]
\caption{LRS2-B VPH Grism Properties} 
\label{tab:GrismProp}
\begin{center}       
\begin{tabular}{|r|c|c|l|} 
\hline
\rule[-1ex]{0pt}{3.0ex} \small{} & \small{\textbf{UV Arm}} & \small{\textbf{Orange Arm}} & \small{Units / Comments} \\
\hline
\rule[-1ex]{0pt}{2.0ex}  \scriptsize{$\lambda_{\mathrm{min}}$, $\lambda_{\mathrm{max}}$} & \scriptsize{370 , 470} & \scriptsize{460 , 700} & \scriptsize{nm} \\
\hline
\rule[-1ex]{0pt}{2.0ex}  \scriptsize{Dispersion} & \scriptsize{0.484} & \scriptsize{1.163} & \scriptsize{\AA\ pixel$^{-1}$ ($1\times1$ binning)} \\
\hline
\rule[-1ex]{0pt}{2.0ex}  \scriptsize{Spectral Resolution} & \scriptsize{2.216} & \scriptsize{5.084} & \scriptsize{\AA\ (FWHM)} \\
\hline
\rule[-1ex]{0pt}{2.0ex}  \scriptsize{Fringe Frequency} & \scriptsize{1765} & \scriptsize{776} & \scriptsize{lines mm$^{-1}$} \\
\hline
\rule[-1ex]{0pt}{2.0ex}  \scriptsize{Bragg Wavelength} & \scriptsize{400} & \scriptsize{590} & \scriptsize{nm} \\
\hline
\rule[-1ex]{0pt}{2.0ex}  \scriptsize{Angle of Incidence} & \scriptsize{11.5} & \scriptsize{6.8} & \scriptsize{Degrees (on VPH Layer)} \\
\hline
\rule[-1ex]{0pt}{2.0ex}  \scriptsize{Fringe Tilt} & \scriptsize{2.0} & \scriptsize{2.0} & \scriptsize{Degrees (see Fig. \ref{fig:GrismSchem})} \\
\hline
\rule[-1ex]{0pt}{2.0ex}  \scriptsize{Assembly Physical Tilt} & \scriptsize{14.7} & \scriptsize{14.3} & \scriptsize{Degrees (see Fig. \ref{fig:GrismSchem})} \\
\hline
\rule[-1ex]{0pt}{2.0ex}  \scriptsize{Prism Wedge Angle} & \scriptsize{15.5 , 19.8} & \scriptsize{N/A , 3.5} & \scriptsize{Degrees (Front, Back)} \\
\hline
\rule[-1ex]{0pt}{2.0ex}  \scriptsize{DCG Optical Thickness} & \scriptsize{3.5} & \scriptsize{6.0} & \scriptsize{$\mu$m} \\
\hline
\rule[-1ex]{0pt}{2.0ex}  \scriptsize{DCG Refractive Index Modulation} & \scriptsize{0.06} & \scriptsize{0.048} & \scriptsize{(assumed sinusoidal)} \\
\hline
\end{tabular}
\end{center}
\end{table}

As outlined in Ref. \citenum{Chonis11}, the wavelength coverage for each channel leads to a very good estimate of the fringe frequency for each grism. Using the Bragg condition (e.g., Ref. \citenum{Baldry04}), we can proceed to find the optimal angle of incidence on the grating layer for the maximization of the diffraction efficiency at a wavelength of our choice (i.e., the Bragg wavelength). For the UV Arm, we choose the Bragg wavelength to be $\sim400$ nm, which is blueward of the central wavelength of the UV Arm coverage so as to compensate for the falling CCD quantum efficiency (QE), fiber transmission, atmospheric transmission and telescope throughput towards the near UV. For the Orange Arm, we choose a more central wavelength of $\sim590$ nm. The physical tilt of the grism assembly is limited by the need to avoid imaging the Littrow recombination ghost\cite{Burgh07}. As pointed out in Ref. \citenum{Burgh07}, the way to decouple the Bragg condition from the Littrow configuration is to introduce a tilt to the fringes. Taking into account the fringe tilt, the wedge angle of the front prism (i.e., the prism onto which the collimated beam is first incident) can be set to give the proper angle of incidence on the grating layer. No front prism is necessary for the Orange Arm. Subsequently, the back prism wedge angle is set to properly place the spectrum on the CCD. For both LRS2-B grisms, we use BK7 glass for the prisms and VPH grating substrates. The resulting spectral resolution for each channel can be estimated by rearranging and simplifying Eq. A6 from Ref. \citenum{Baldry04} in terms of the VIRUS instrument geometry. For a fiber core diameter of 170 $\mu$m, we find that the LRS2-B spectral resolution is 2.216 \AA\ (158 km/s FWHM; $R \approx 1895$ at the central wavelength) for the UV Arm, which is sufficient for carrying out LAE follow-up studies. For the Orange Arm, we find a resolution of 5.084 \AA\ (262 km/s FWHM; $R \approx 1140$). In reality, the spectral resolution is slightly better than these calculated values when considering the added dispersion from the back prisms, but the effect is not significant. For the VIRUS CCD (see $\S$\ref{subsec:camera}) binned $1\times1$, the dispersion for the UV and Orange Arms are 0.484 \AA\ pixel$^{-1}$ and 1.163  \AA\ pixel$^{-1}$, respectively. The grisms can be seen schematically in Fig. \ref{fig:GrismSchem} along with a ray-trace for each channel. Additionally, the properties of each channel are tabulated in Table \ref{tab:GrismProp}.

   \begin{figure}[t]
   \begin{center}
   \begin{tabular}{c}
   \includegraphics[width=0.98\textwidth]{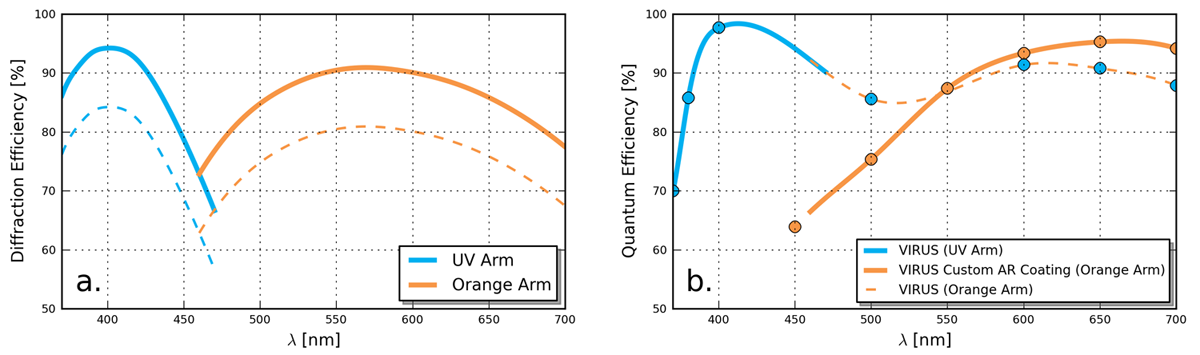}
   \end{tabular}
   \end{center}
   \caption[example] 
   { \label{fig:RCWA-QE}   
\textit{a}) RCWA predictions of the first order external diffraction efficiency of the LRS2-B UV and Orange Arm grisms. The solid curves are the predictions taking into account the internal transmittance of the prism and substrates, reflection losses, and the transmittance of the DCG layer. The dashed curves show the 10\% lower envelope of what we can accept, which is a more realistic expectation of the final product. The parameters describing each model are shown in Table \ref{tab:GrismProp}.\\ \textit{b}) The QE of the LRS2-B CCD detectors. The data points are actual measurements of the QE while the curves are cubic spline fit to the data. Blue corresponds to the VIRUS CCD that is will be used for the LRS2-B UV Arm. Orange corresponds to the Orange Arm CCD detector, which is a standard VIRUS CCD with a custom AR coating to boost the efficiency towards 700 nm. The dashed orange curve is the extension of the standard VIRUS CCD QE spline fit to show how it would perform if used in the Orange Arm without the custom coating. 
}
   \end{figure} 

Using the valuable reviews\cite{Barden00,Baldry04,Burgh07} of VPH grating physics in the literature, we have explored the potential diffraction efficiencies that our grisms may achieve which will be useful in later estimating the instrument's performance (see $\S$\ref{subsec:performance}). In Fig. \ref{fig:RCWA-QE}$a$, we present predictions of the first order external diffraction efficiency of each grism calculated using Rigorous Coupled Wave Analysis (RCWA\cite{Gaylord85}) for unpolarized light. These theoretical diffraction efficiencies have been corrected to include the internal transmittance of the BK7\footnote{We have assumed Schott N-BK7HT internal transmittance:\\ \textit{http://edit.schott.com/advanced\_optics/us/abbe\_datasheets/schott\_datasheet\_n-bk7ht.pdf}} prisms and substrates, a 1.2\% average loss due to surface reflections with an average anti-reflection (AR) coating, and the transmittance of the dichromated gelatin (DCG) diffracting layer (as scaled from the data in Ref. \citenum{Barden00}). We can accept a grism with diffraction efficiency no more than 10\% worse than these predictions. Our extensive experience with the VPH-based dispersing elements has shown that meeting the lower limit of this requirement is possible\cite{Hill03,Adams08,Chonis12b}. 

As can be seen in Fig. \ref{fig:GrismSchem}, the grisms have a square footprint with 150 mm sides to ease their mounting in the instrument (see the following section). However, the grating layer clear aperture is circular with an over-sized 140 mm diameter as compared to the 125 mm diameter collimated beam to relax the translational position tolerance of the grism. We have performed a detailed Monte Carlo analysis of the grism tolerance, which includes a tolerance of the individual components, the assembly of those components, and the position of the assembly in the instrument. Our criteria for determining an acceptable tolerance was that no fiber images can fall off the CCD (which constrains the image motion to $\pm110$ $\mu$m in the spatial direction) and that no more than 45\% of the nominal 10 nm transition region between the two channels' spectral coverage could be infringed upon. From this analysis, we have determined that each grating is straightforward to fabricate. Positioning and aligning the grism in the collimator assembly is also straightforward (see the following section): the translational tolerance is trivial and the only rotational alignment that requires fine adjustment as a compensator is that about the camera's optical axis (see Fig. \ref{fig:GrismSchem}). From 500 Monte Carlo realizations of the grisms, we find that the range of rotational compensation about the camera's optical axis for the UV and Orange Arm is $\pm0.4$\degree\ and $\pm0.6$\degree, respectively. The adjustment must be able to be controlled within 0.1\degree. In Fig. \ref{fig:DichroicCross}, we show the effect of the fabrication and alignment tolerance on the position of the edges of each channel's spectral coverage in the transition region. These variations are dominated by the $\pm2$ line mm$^{-1}$ variation of the grating layer fringe frequency, which is currently at the limit of VPH technology.

\subsection{Collimator Assembly}\label{subsec:collimator}
   \begin{figure}[t]
   \begin{center}
   \begin{tabular}{c}
   \includegraphics[width=0.95\textwidth]{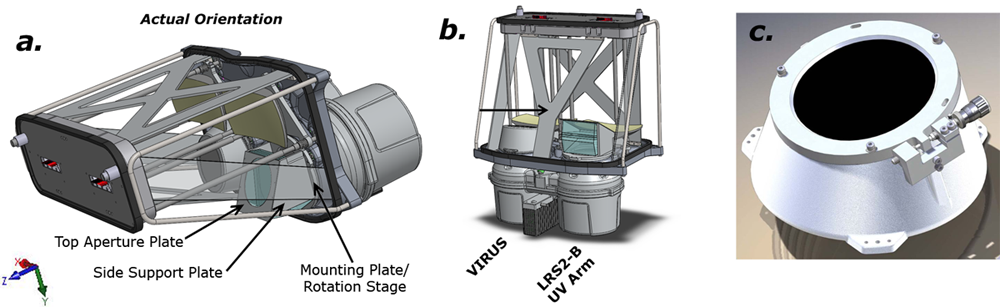}
   \end{tabular}
   \end{center}
   \caption[example] 
   { \label{fig:CollimatorAssembly}   
LRS2-B collimator assembly. \textit{a}) A rendering of a spectrograph pair. If looking at the instrument from the fiber ``slit'' end, the right channel is the LRS2-B UV Arm while the left channel is VIRUS for comparison. The rendering is shown in the actual orientation of the instrument in its rack mount. The major components of the grism mount are labeled. In the lower left corner, we show the coordinate system to which the text of $\S$\ref{subsec:collimator} refers. The origin of this coordinate system is intended to coincide with the center of the grism's grating layer, but is not shown as such in the figure for visual clarity. \textit{b}) A bottom view of the spectrograph pair shown in $a$. The ``Y'' support strut that avoids a collision with the larger grisms and is custom for the LRS2-B collimator assembly is arrowed. \textit{c}) A rendering of the VIRUS grating mount, showing the adjustable top cell that is manually actuated by a micrometer and opposing spring plunger. A similar rotation mechanism will be built into the grism's mounting plate (see $a$) to allow rotation of the grism about the camera's optical axis for fine adjustment. 
}
   \end{figure} 
Other than the change from the standard VPH gratings used in VIRUS to the larger VPH grisms used in LRS2-B, the LRS2-B collimator assembly mostly remains optically and mechanically unchanged from the standard VIRUS collimator assembly. As seen in the previous section, the LRS2-B grisms are physically a different shape and size from the VIRUS gratings. For the UV Arm, the grism is large enough to collide with a support strut on the lower side of the standard VIRUS collimator assembly. This ``Y'' support strut is only for stiffness, and can easily be modified to accommodate the larger optic, as seen in Fig. \ref{fig:CollimatorAssembly}$b$. 

The only additional change to the standard VIRUS collimator assembly for LRS2-B is custom mounting cells for the grism dispersers for both channels, which are currently at a conceptual design stage. As mentioned in the previous section, our Monte Carlo tolerance analysis for both grism fabrication and alignment in the instrument showed that the only fine adjustment necessary for aligning the grism is for the rotation of the assembly about the camera optical axis (the $Z$ axis as shown and described in Fig. \ref{fig:CollimatorAssembly}$a$). This rotation requires fine adjustment since it serves as a compensator for angular misalignments in the grism assembly (such as the fringe direction or alignment errors of the prisms relative the fringes) that would otherwise result in the movement of fibers (or sections of fibers' spectra) off the CCD detector. For both LRS2-B channels, rotations about the $X$ axis only affect the angle of incidence (which controls the diffraction efficiency blaze function) and are constrained to $\pm0.25$\degree. Rotations about the $Y$ axis affect the position and tilt of the spectra on the CCD, which when constrained to $\pm0.25$\degree\ can be compensated for by the $Z$ axis rotation. The $\pm0.25$\degree\ tolerance on the $X$ and $Y$ rotations over the 75 mm half-length of the grism corresponds to $\pm0.3$ mm, which can easily be met by machine accuracy. The translational tolerance of the grism assembly for both channels is also easy to meet as the center of the grism must be within $\pm2$ mm of the nominal position along each of the $X$, $Y$, and $Z$ axes thanks to the over-sized grating clear aperture.

For the grism mounts, we adopt the same semi-permanent method as VIRUS of utilizing RTV adhesive to epoxy the dispersing optic into a cell (see Fig. \ref{fig:CollimatorAssembly}$a$ and $b$). The cell will be constructed with aluminum side support plates that are mounted onto a custom rotation stage which attaches to the collimator assembly's base plate using the same bolt and locating pin pattern as used by the VIRUS grating mounts. The latter allows the use of a collimator base plate that is fabricated in production with those for VIRUS. The side support plates will be machined with heights and angles specific to each spectrograph channel's grism geometry and setup. Mounted on top of the side support plates will be a top plate that contains an aperture stop for the grism onto which the front prism face will be epoxied. A machined pocket on the underside of the top aperture plate will ensure the placement of the grism in the cell to within the $\pm2$ mm required accuracy. RTV will also be injected into the mount to form a bond between the sides of the grism and the inside face of the side support plates. Removable supports machined into the side support plates will hold the grism securely in place while the epoxy cures. The rotation stage on which the side support plates are mounted will have a similar design to the VIRUS grating mounts\cite{Prochaska12}. In Fig. \ref{fig:CollimatorAssembly}$c$, we show the VIRUS grating mount rotation stage for reference. In short for LRS2-B, the grism cell mounting plate will contain two pins that will slide in arced slots on the rotation stage. A micrometer and opposing spring plunger will push on a machined tab on the rotation stage plate to rotate the cell about the $Z$ axis. Once the spectra are aligned on the CCD detector, a series of locking bolts can be tightened to lock the rotation stage to the mounting plate. The design of the grism mounting cell is fairly simple, and we are ready to proceed with a more concrete design that will be completed while we are waiting for the grisms to be delivered.

   \begin{figure}[b]
   \begin{center}
   \begin{tabular}{c}
   \includegraphics[width=0.95\textwidth]{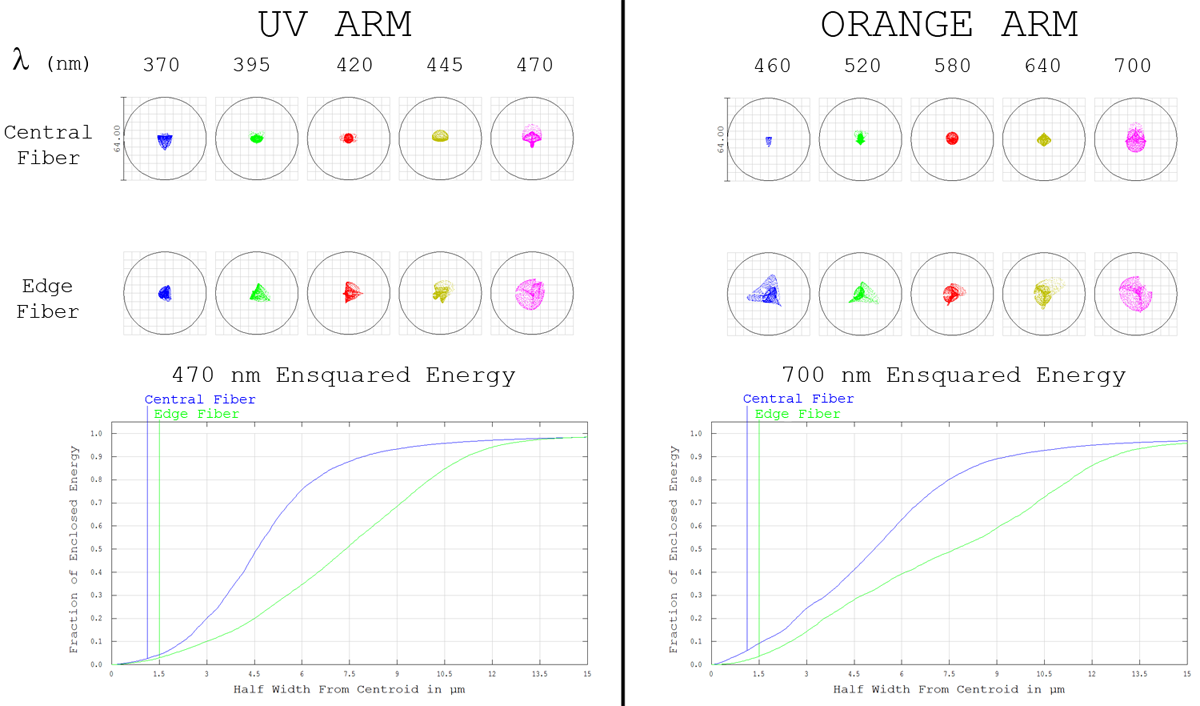}
   \end{tabular}
   \end{center}
   \caption[example] 
   { \label{fig:ImageQuality}   
The focal plane image quality for the LRS2-B UV Arm (left column) and Orange Arm (right column). We show the spot diagrams for five representative wavelengths for each spectrograph channel for both the central fiber of the fiber ``slit'' and one of the edge fibers to show the full range of image quality. The circle around each spot is 64 $\mu$m in diameter, which is the size of the reimaged fiber core. For both spectrograph channels, the poorest image quality is seen in the edge fiber at the reddest wavelength (which is imaged at the corner of the CCD detector). To show that even the worst image quality meets the EE90 within 2 pixels requirement, we show the diffraction ensquared energy within a $2\times2$ pixel area for the reddest wavelength of each channel in each of the two example fibers. 
}
   \end{figure} 

\subsection{Cameras and Detector System}\label{subsec:camera}
Since the spectral coverage is completely within that of VIRUS for the UV Arm, no changes to the standard VIRUS camera design are required. For the Orange Arm, only minor modifications are required since the spectral coverage is extended 150 nm redward of the nominal VIRUS design. The main optical change is that a custom fused silica aspheric field flattener is necessary. With the standard VIRUS ashperic field flattener, the Orange Arm image quality for wavelengths $>680$ nm in the 20\% of fibers farthest from the optical axis drops below the requirement, which is $\geq 90$\% diffraction ensquared energy (EE90) within 2 pixels for all wavelengths in each fiber. Although the fiber spacing on the CCD is less aggressive than VIRUS, meeting this image quality specification is important to avoid any fiber cross-talk. This is increasingly important for the redder wavelengths where bright night sky emission lines begin to become a factor. The Orange Arm field flattener optimization is simple and only requires a different even aspheric surface on the rear side of the optic and a slightly smaller thickness. We are able to retain the same radius of curvature for the front surface as well as the same distance as for VIRUS from the CCD detector to the vertex of the back aspheric surface, which allows the use of the standard VIRUS mounting hardware for the custom optic. In Fig. \ref{fig:ImageQuality}, we show the focal plane image quality for each of the two spectrograph channels, which achieves the EE90 requirement even at the very corners of the CCD detector. In addition to the custom field flattener prescription, the Orange Arm transmissive optics all have custom coatings to extend the wavelength coverage with maximum throughput to 700 nm. As with the collimator reflective optics, the standard VIRUS camera primary mirror can be used for the Orange Arm. 

The LRS2-B cameras utilize the same $2048\times2048$ CCD detectors as VIRUS. The detectors are thinned UV-blue optimized devices with 15 $\mu$m square pixels. The UV Arm CCD is simply a standard VIRUS CCD. For the Orange Arm, we have the option of using a standard VIRUS CCD or utilizing a custom AR coating to help boost the QE for wavelengths $> 600$ nm. In Fig \ref{fig:RCWA-QE}$b$, we show the measured QE of the two LRS2-B CCDs that we have procured (see $\S$\ref{subsec:procurement}). The CCDs feature high QE (especially in the UV-blue for the UV Arm CCD), and low read-noise of $\leq3.1$ e$^{-}$. The detector system is being supplied by Astronomical Research Cameras, Inc.\cite{Tuttle12}, and can read a $2\times1$ binned CCD in $\sim20$ seconds. 

By using the same cameras and detector system as VIRUS, we are able to utilize the design and investment made in its cryostat design and its infrastructure for cryogenics\cite{Chonis10} and CCD readout\cite{Tuttle12}. LRS2-B will fit seamlessly into the VIRUS support structure on the telescope and its cryogenic camera pair will be connected to the extensive VIRUS liquid nitrogen distribution system via the novel make-break bayonet thermal connector\cite{Chonis10}. For VIRUS, periodic warm-up and re-pumping of the camera is acceptable and expected since all spectrographs will not be on-line simultaneously. Unlike the VIRUS cameras, the LRS2-B camera pair will be outfitted with full-range vacuum gauges and ion pumps to ensure the longevity of the vacuum since we wish to avoid down time. Experience with the current LRS indicates that the vacuum can be maintained for $>18$ months as long as the instrument remains cold. 

\subsection{Predicted Performance}\label{subsec:performance}
We can make reliable predictions of the performance of LRS2-B based on experience and performance of the VIRUS prototype, the Mitchell Spectrograph\cite{Hill08b}, whose performance exceeds that required for HETDEX. With typical shifts of the image of $\sim1/20$ of a resolution element over a 5\degree\ Celsius temperature change, the Mitchell Spectrograph has proven to be extremely stable with temperature. We expect LRS2-B to perform at least this well when considering improvements to the VIRUS camera design. Additionally, we note that the instrument is rack mounted in the VIRUS support structure and is thus invariant to gravity induced flexure. The image quality of the instrument (see Fig. \ref{fig:ImageQuality}) is expected to be excellent and should prove beneficial for the red end of LRS2-B's spectral coverage where bright night sky emission lines become more numerous. Through a detailed Monte Carlo analysis\cite{Lee10b}, the design of VIRUS is such that a 1.0 pixel variation in the FWHM point spread function (PSF) results from the manufacturing and alignment tolerance of any interchangeable collimator and camera pair. As a one-off instrument, considerably more time will be spent in the alignment of LRS2-B to meet the image quality requirements for the higher resolution of the instrument. We expect to meet such requirements since we have already proven the concept for the higher resolution modes of the Mitchell Spectrograph.

\begin{figure}[t]
   \begin{center}
   \begin{tabular}{c}
   \includegraphics[width=0.98\textwidth]{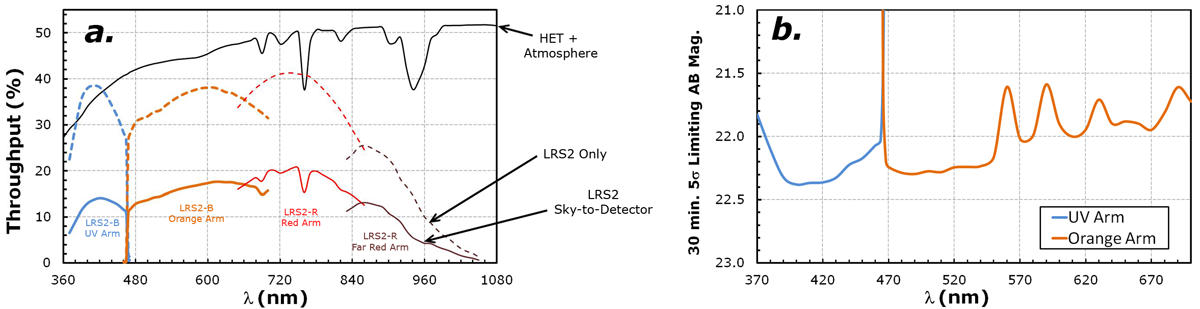}
   \end{tabular}
   \end{center}
   \caption[example] 
   { \label{fig:Performance}   
\textit{a}) The predicted throughput of LRS2. The black curve shows the total throughput of the upgraded HET, including the atmosphere at an airmass of 1.22. The colored dashed curves show the throughput of LRS2 for each channel as estimated from the fiber feed to the detector. The solid colored curves show the final sky-to-detector throughput of the entire system. As labeled, the thick curves represent the two LRS2-B channels, while the thinner curves represent LRS2-R (see $\S$\ref{sec:lrs2-r}). Note that unlike the LRS2-B channels, the LRS2-R channels do not include the transmittance/reflectance of the dichroic since it has not yet been designed. \textit{b}) The 5$\sigma$ detection limit (given in AB magnitudes) in a 30 minute exposure for the two LRS2-B channels (see the text for details).
}
   \end{figure}

As first shown in Ref. \citenum{Lee10}, we present an updated model of the throughput of LRS2-B, which is based on the model used for VIRUS that has been verified against on-sky tests\cite{Hill08b} by the Mitchell Spectrograph. For LRS2-B, additional factors taking into account the lenslet coupling and feed optics, the dichroic, the obstruction of the detector, the smaller resolution element in wavelength space, as well as a more realistic estimate of the grism external diffraction efficiency and the measured detector QE. Fig. \ref{fig:Performance}$a$ shows the throughput for each spectrograph channel as a function of wavelength. Due to the integral field feed of LRS2-B, there are no aperture losses included in the model (as would be the case for a slit spectrograph). Fig. \ref{fig:Performance}$b$ shows the 5$\sigma$ detection limit (in AB magnitudes) for a 30 minute exposure for each of the two LRS2-B channels, assuming clear and dark sky conditions. The sensitivity predictions for the two channels assume 1.2\arcsec\ FWHM seeing for an unresolved source and that the PSF is perfectly centered on an IFU spaxel. Due to the finely sampled IFU layout relative to the HET median image quality, the sensitivity curves presented in Fig. \ref{fig:Performance}$b$ vary as functions of the seeing (i.e., fainter detection limits can be achieved with improved seeing) as well as PSF location relative to the IFU spaxel grid (i.e., fainter detection limits can be achieved if the source is perfectly centered on an IFU lenslet).

\subsection{Operation}\label{subsec:operation}
The operation of LRS2-B will be through the HET queue scheduling system. Since LRS2-B is based on VIRUS, its operation will be congruent with the VIRUS spectrograph array. LRS2-B will be rack mounted in the VIRUS support structure\cite{Heisler10} and will connect to the large liquid nitrogen cryogenic plumbing manifold that provides coolant for the VIRUS spectrograph array\cite{Chonis10}. While the two LRS2-B CCDs share a common data system architecture with VIRUS, they will be read out through a dedicated computer for data acquisition rather than through the multiplexed VIRUS detector system since the instruments are operated independently. However, the data acquisition systems for LRS2-B and VIRUS must be able to communicate with one another since they share the same shutter (see below), and the CCD clocks will be synchronized to minimize noise. 

The main operational change between LRS2-B and the current LRS is from slit to IFU spectroscopy. The IFU, in addition to the improved metrology\cite{Lee12} provided by the HET WFU, should reduce target acquisition time significantly to around one minute from the current value of $\sim10$ minutes. The IFU is also a necessity for observing with LRS2-B and VIRUS together in parallel mode\cite{Odewahn12}. This is because three dithers of $\sim1$\arcsec\ are required to fill in the field of a standard VIRUS observation, resulting in offsets of the image at the LRS2-B input. As such, the read out of LRS2-B and VIRUS must be coordinated and will provide the natural ability to obtain three independent exposures for each LRS2-B observation. Additionally, the relative offsets are extremely precise\cite{Vattiat12} so that the movements can be easily communicated back to the user for later use during data reduction. The data reduction software for LRS2-B is a simple reconfiguration of the CURE software pipeline that is already written to reduce VIRUS data. This software, as well as that for reducing data from the Mitchell Spectrograph\cite{Adams11}, have demonstrated Poisson noise dominated sky subtraction over $350<\lambda(\mathrm{nm})<585$ for hundreds of nights of observations, and tests have been successful extending out to 670 nm.

\section{STATUS}\label{sec:status}
\subsection{Procurement}\label{subsec:procurement}
\begin{figure}[t]
   \begin{center}
   \begin{tabular}{c}
   \includegraphics[width=0.98\textwidth]{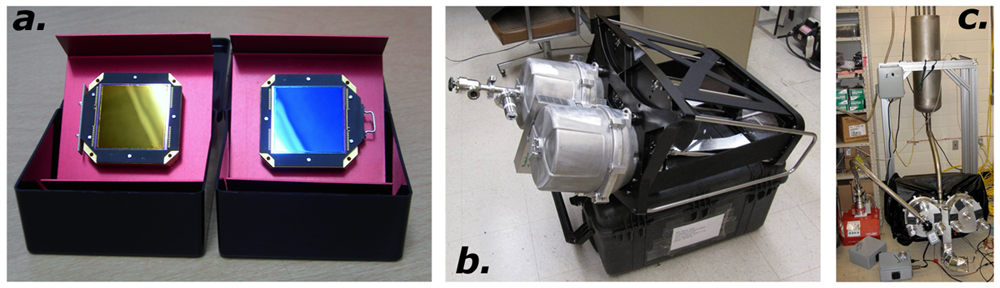}
   \end{tabular}
   \end{center}
   \caption[example] 
   { \label{fig:ProcureConstruct}   
\textit{a}) The LRS2-B CCD detectors. The CCD at left (reflecting yellow) is a standard VIRUS CCD with particularly good characteristics that will be used for the UV Arm. The CCD at right (reflecting blue) is a VIRUS CCD with a custom AR coating that is intended for the Orange Arm. The custom coating helps to boost the QE at longer wavelengths.\\ \textit{b}) One of the first VIRUS unit spectrograph pairs off the production line. \textit{c}) The first VIRUS unit spectrograph pair under testing. The tank above the instrument is a liquid nitrogen dewar that is cooling the camera using the same make-break bayonet thermal connector\cite{Chonis10} that will be used on the up-scaled and multiplexed cryogenic manifold on the telescope. The large knobs on the back of the camera are temporary and allow for the alignment of the camera primary mirror while under vacuum. 
}
   \end{figure}

Since the majority of LRS2-B's optical and mechanical components are identical to VIRUS, they have already been procured in the VIRUS production lines from a number of vendors\cite{Hill12a,Tuttle12}. For the collimator assembly, the only components that have not yet been procured for LRS2-B specifically are the custom lower ``Y'' support, the grism mounts, and the grisms themselves. The grisms are the longest lead-time items for the collimator, but are ready for order. The grism mounting cells are straightforward to design, and will be procured while we are waiting for the grisms to be delivered. For the camera, all components for both channels are on-hand and we are ready to begin construction this summer. In Fig. \ref{fig:ProcureConstruct}$a$, we show the two LRS2-B CCDs (note that their measured QE is shown in Fig. \ref{fig:RCWA-QE}$b$). The coating for the Orange Arm CCD appears to be optimized too red, so it is possible that another VIRUS CCD with especially good QE may be repurposed for the LRS2-B Orange Arm since the measured QE for the VIRUS CCD utilized in the UV Arm is $>85$\% out to 700 nm. For the IFU and feed optics, we have already ordered the dichroic and are ready to order to microlens array, but the remainder of the feed assembly is not yet ready since we are finalizing the design in order to fully understand the fabrication and alignment error budget. We expect this to be complete by the end of the summer. The IFU will also be ordered on that time scale, and we expect no issues with fabricating it since we have ample experience with assembling VIRUS IFUs. We expect to have a fully constructed instrument that is ready for lab testing by the year's end in anticipation for first light on the HET during quarter 4 of 2013. 

\subsection{Construction and Lab Tests}\label{subsec:construction}
While we have not yet constructed the collimator and camera that are specifically going to be paired together and used for LRS2-B, we have already assembled several complete VIRUS spectrographs and have completed successful labs tests. A constructed VIRUS pair can be seen in Fig. \ref{fig:ProcureConstruct}$b$ and a second VIRUS pair in the process of being tested in the lab is shown in Fig. \ref{fig:ProcureConstruct}$c$. Ref. \citenum{Tuttle12} discusses the first results and performance of the first VIRUS spectrographs that were assembled on the production line. Since we have already assembled multiple VIRUS pairs, we expect the construction of LRS2-B to proceed smoothly. Extra time and care will be spent in the alignment of the instrument since its camera pair and collimator assembly will permanently remain coupled (unlike a typical VIRUS spectrograph pair that is required to maintain its image quality throughout interchangeability with any other camera pair or collimator\cite{Lee10b}). The largest challenge in the assembly of LRS2-B will be the alignment of the IFU feed optics. While the alignment of the spectrograph itself is well understood\cite{Tuttle12}, we are currently working to understand how to construct and properly align the LRS2-B specific IFU feed optics to maintain the required image quality for feeding the IFU fibers. 

\section{LRS2-R: EXTENDING REDWARD}\label{sec:lrs2-r}
Although the original LRS2 concept of a quad-channel instrument simultaneously covering $350<\lambda(\mathrm{nm})<1000$ is no longer a viable option, the extension of LRS2-B's capabilities redward has garnered considerable interest from the HET community. LRS2-B and LRS2-R will be highly complementary to each other and the addition of LRS2-R to extend the wavelength coverage (albeit not simultaneously) to 1 $\mu$m will provide the largest improvements in sensitivity and capability over the current LRS. Below, we very briefly discuss the science motivation and instrument design of LRS2-R.

\subsection{Science Motivation}\label{subsec:lrs2-rscience}
The main advantage of LRS2-R is that it will extend the scientific capabilities of LRS2-B to higher wavelength (and thus higher redshift; see the discussion of science topics in $\S$\ref{sec:science}), and will enable a large number of science programs that are not currently viable or are difficult with LRS due to the increased sensitivity (see the following section). As with LRS2-B, LRS2-R will especially play a unique and important role as a follow-up instrument for large surveys (especially for the large sample of objects discovered in HETDEX) due to its high efficiency and broad spectral coverage as well as excellent sky subtraction capabilities. 

As an example, the Dark Energy Survey (DES; Ref. \citenum{DES}) will image many fields that will be accessible to the HET (such as frequent observations of several equatorial fields for a supernova survey and SDSS Stripe 82) to $\sim24$ mag in $grizY$ bands. The DES Supernova Survey will discover and measure light curves for $\sim$3000 Type Ia supernovae in 5 years for $0.3<z<1.0$. LRS2-R could be used for follow-up spectroscopy of a large number of Type Ia candidates as well as their host galaxies to confirm redshifts. The predicted sensitivity of LRS2-R ($\sim23$ mag for a 5$\sigma$ detection in 30 minutes; see Ref. \citenum{Lee10}) is also well matched to the expected discovery brightness of supernovae in DES fields. DES will identify a large number of high redshift QSO candidates that will require spectroscopic follow-up, especially for $z>6$ (which will need spectroscopic confirmation since photometric selection methods are not always reliable at such high redshift) and for the confirmation and calibration of photo-$z$ measurements for $z>4$ candidates that could be used for clustering studies. DES will also discover many galaxy clusters over a moderate range of redshifts that will be used for cluster-cosmology science and studies of galaxy evolution. LRS2-R could be used to survey candidate clusters and at a minimum could provide validation of the cluster-finding algorithms and assess the completeness of galaxy catalogs.

LRS2-R will also be a powerful instrument for follow-up studies and redshift confirmation of galaxies at high redshift. For LRS2-B, the highest redshift for which the \lya\ transition is observable is $z\approx4.7$. For LRS2-R, this is extended to $z\approx7.5$, which when coupled with a 10 m class telescope such as the HET can yield a powerful combination for studying the early universe. This provides the opportunity to study high redshift quasars and other AGN as well as probing reionization using \lya\ emitting galaxies. The fraction of \lya\ emitters appears to increase with increasing redshift to $z\approx6$, as might be expected for the less evolved galaxies observed at earlier cosmic times\cite{Stark10}. While the number of candidate $z\gtrsim7$ galaxies is increasing from photometrically selected samples, spectroscopic confirmation is scarce. LRS2-R could follow-up new $z\approx7$ galaxy candidates to confirm redshifts using \lya. While the total throughput of LRS2-R drops for the relevant wavelengths (see Fig. \ref{fig:Performance}), it has the advantage of the unity fill-factor IFU so that slit losses are nonexistent. Recent observations suggest that the fraction of \lya\ emitters drops near $z\approx7$, which could indicate an increase in the neutral fraction of hydrogen at that epoch\cite{Stark10}. An LRS2-R survey to follow-up on candidate $z\approx7$ galaxies to confirm the tentative results that currently exist with the limited spectroscopically confirmed galaxies would place important constraints on the epoch of reionization. 

Spectroscopy from 600 nm to 1$\mu$m is critical for identifying and studying brown dwarfs since the M and L spectral types are identified by molecular absorption bands in the red\cite{Kirkpatrick99}, \ha\ can be used as a diagnostic of accretion from circumstellar disks\cite{Muzerolle05}, and gravity sensitive absorption lines can place constraints on masses\cite{Cruz07}. A use of LRS in this field has been for the confirmation of brown dwarf candidates identified in photometric surveys of young stellar objects in star-forming regions\cite{Luhman09}. With the increased sensitivity of LRS2-R over the current LRS at the relevant wavelengths, surveys for these young stellar objects with the HET could reach mass limits as low as $\sim$5 $M_{\mathrm{Jupiter}}$ for L spectral types with $I<22$ mag. This is important as measuring the minimum mass at which stellar-like objects can form can provide fundamental tests of the theories of star and planet formation. 

\subsection{Instrument Design}\label{subsec:lrs2-rdesign}
The LRS2-R instrument design concept is identical to that of LRS2-B. Although not simultaneously, LRS2-B and LRS2-R together will provide the broad and efficient spectral coverage envisaged in the original quad-channel LRS2 design\cite{Lee10}. The wavelength coverage extension to 1 $\mu$m requires more extensive modification to the nominal VIRUS design than for LRS2-B, especially for the camera assembly which will utilize new thick deep-depletion CCD detectors for increased red efficiency. LRS2-R will be a stand-alone instrument, and will have its own set of feed optics. Other than coatings and a custom dichroic beam splitter (with a transition around 830 nm), the LRS2-R feed optics and IFU are identical to LRS2-B (see $\S$\ref{subsec:IFUfeed}). The LRS2-R IFU feed will reside next to LRS2-B's at the center of the HET focal plane, enabling parallel mode observations with VIRUS and the quick transition of targets between LRS2-R and LRS2-B when the full wavelength coverage from 370 nm to 1 $\mu$m is desired. The collimator assembly is also optically and mechanically identical to LRS2-B, except that all optics have custom red-optimized coatings (see $\S$\ref{subsec:grisms} and \ref{subsec:collimator}). LRS2-R also must use VPH grisms within the fixed VIRUS mechanical framework. The grisms will be designed so that the instrument has $R\approx1800$ for both spectrograph channels (for LRS2-R, the channels are called the Red and Far-Red Arms, covering $650<\lambda(\mathrm{nm})<835$ and $825<\lambda(\mathrm{nm})<1050$, respectively) to ensure good sky subtraction in the far red where strong OH bands dominate the sky. Despite the lower spectral resolution, the LRS2-R channels will have grisms with comparable prism angles to the LRS2-B UV Arm grism due to the increased beam deviation necessary for the redder wavelengths. 

The most extensive modifications to the nominal VIRUS design for LRS2-R are for the camera and detector system. LRS2-R will utilize thick deep-depletion E2V CCDs with the same $2048\times2048$, 15 $\mu$m pixel format as VIRUS. The CCDs will be AR coated to optimize their red response and to minimize fringing in the Far-Red Arm. The CCD controllers have the same requirements as VIRUS and LRS2-B (i.e., low read noise of $<4$ e$^{-}$ and fast read out) and will also be supplied by Astronomical Research Cameras, Inc. Only slight modifications to the standard VIRUS CCD controller are required, so the LRS2-R detectors can be read out by the same data system architecture. As with LRS2-B, LRS2-R will have its own dedicated computer for independent data acquisition. Mechanically, the camera assembly is the same as for VIRUS and LRS2-B except for the detector head to mount the CCDs, which have a different package design. Modification to the invar ``spider'' that suspends the detector head at the internal focus of the Schmidt camera will be required in addition to the mounting scheme for the field flattener lens. Optically, the LRS2-R camera is conceptually the same as VIRUS and LRS2-B. The camera mirror will have a red-optimized reflective coating and the transmissive optics (i.e., the Schmidt corrector plate and field flattener) will have custom AR coatings. We have performed an image quality analysis for LRS2-R and have discovered the need to reoptimize the camera's transmissive optics for the longer wavelengths to meet the EE90 requirement. This is especially important for LRS2-R for minimizing cross-talk between fibers where the night sky emission is more prominent. We have found an optimization solution that gives excellent image quality for both the Red and Far-Red Arms simultaneously. The design includes a custom field flattener as well as a custom corrector plate, both of which are fabricated from fused silica with slightly different thickness, aspheric surface prescriptions, and radii of curvature (for the field flattener) from the nominal VIRUS values.

LRS2-R will provide significant improvement in the sensitivity over the current LRS at comparable wavelengths. This is largely due to the increased sensitivity of the CCDs and the greatly reduced fringing. In Fig. \ref{fig:Performance}, we have included the preliminary predicted throughput of the LRS2-R channels. For LRS2-R, this is calculated with preliminary estimates of the detector QE, grism diffraction efficiency, and an estimation of the average dichroic efficiency (although we have not modeled the cross-over region). The addition of LRS2-R in combination with LRS2-B will provide the HET community with competitive and efficient low resolution spectroscopy from the near UV to the NIR. 

\section{OUTLOOK}\label{sec:outlook}
As has been described in this paper, LRS2 will be a worthy replacement for the workhorse LRS instrument for the upgraded HET. With the new spectroscopic and imaging capabilities that result from the other instruments included in the WFU\cite{Vattiat12}, LRS2 will add integral field capability to and greatly improve upon the efficiency of observing single objects for follow-up studies, which has historically been the most widely used mode of LRS. LRS2-B is currently being built and will be ready for commissioning on the HET during quarter 4 of 2013. If there are any significant delays in the HET WFU or if the instrument is completed early, LRS2-B can be readily commissioned on another available telescope, such as the McDonald Observatory 2.7 m. LRS2-R is currently still in a design phase and is awaiting final funding. Since its concept is highly regarded by the HET community, we expect it to be complete within a year of the completion of LRS2-B. The enhanced red sensitivity of LRS2-R over LRS will allow the HET to remain competitive up to the NIR and enable a large number of new science topics ranging from brown dwarfs to high redshift galaxies.    

\acknowledgments     
HETDEX is run by the University of Texas at Austin McDonald Observatory and Department of Astronomy with participation from the Ludwig-Maximilians-Universit\"{a}t M\"{u}nchen, Max-Planck-Institut f\"{u}r Extraterrestriche-Physik (MPE), Leibniz-Institut f\"{u}r Astrophysik Potsdam (AIP), Texas A\&M University, Pennsylvania State University, Institut f\"{u}r Astrophysik G\"{o}ttingen, University of Oxford, and Max-Planck-Institut f\"{u}r Astrophysik (MPA).  In addition to Institutional support, HETDEX is funded by the National Science Foundation (grant AST-0926815), the State of Texas, the US Air Force (AFRL FA9451-04-2-0355), and generous support from private individuals and foundations.

\bibliography{ms_arXiv}         

\begin{thebibliography}{10}

\bibitem{Chonis14}
Chonis, T.~S., Hill, G.~J., Lee, H., Tuttle, S.~E., and Vattiat, B.~L.,
  ``{LRS2}: the new facility low resolution integral field spectrograph for the
  {H}obby {E}berly {T}elescope,'' in [{\em Ground-based and Airborne
  Instrumentation for Astronomy V}{\nolinebreak\hspace{0.1em}]},  {\em Proc.
  SPIE} {\bf 9147-9} (2014).

\bibitem{Hill12b}
Hill, G.~J. et~al., ``Current status of the {H}obby-{E}berly {T}elescope wide
  field upgrade,'' in [{\em Ground-based and Airborne Telescopes
  IV}{\nolinebreak\hspace{0.1em}]},  {\em Proc. SPIE} {\bf 8444-19} (2012).

\bibitem{Hill08a}
Hill, G.~J. et~al., ``The {H}obby-{E}berly {T}elescope {D}ark {E}nergy
  {E}xperiment ({HETDEX}): {D}escription and {E}arly {P}ilot {S}urvey
  {R}esults,'' in [{\em Panoramic Views of Galaxy Formation and
  Evolution}{\nolinebreak\hspace{0.1em}]},  {\em ASP Conference Ser.} {\bf 399,
  115} (2008).

\bibitem{Hill12a}
Hill, G.~J. et~al., ``{VIRUS}: production of a massively replicated fiber
  integral field spectrograph for the upgraded {H}obby-{E}berly {T}elescope,''
  in [{\em Ground-based and Airborne Instrumentation for Astronomy
  IV}{\nolinebreak\hspace{0.1em}]},  {\em Proc. SPIE} {\bf 8446-21} (2012).

\bibitem{Hill98}
Hill, G.~J., Nicklas, H.~E., MacQueen, P.~J., Tejada, C., Duenas, F. J.~C., and
  Mitsch, W., ``The {H}obby-{E}berly {T}elescope low-resolution spectrograph,''
  in [{\em Optical Astronomical Instrumentation}{\nolinebreak\hspace{0.1em}]},
  {\em Proc. SPIE} {\bf 3355-375} (1998).

\bibitem{Lee10}
Lee, H., Chonis, T.~S., Hill, G.~J., DePoy, D.~L., Marshall, J.~L., and
  Vattiat, B.~L., ``{LRS2}: a new low-resolution spectrograph for the
  {H}obby-{E}berly {T}elescope,'' in [{\em Ground-based and Airborne
  Instrumentation for Astronomy III}{\nolinebreak\hspace{0.1em}]},  {\em Proc.
  SPIE} {\bf 7735-276} (2010).

\bibitem{Vattiat12}
Vattiat, B.~L. et~al., ``Design, testing, and performance of the
  {H}obby-{E}berly {T}elescope prime focus instrument package,'' in [{\em
  Ground-based and Airborne Instrumentation for Astronomy
  IV}{\nolinebreak\hspace{0.1em}]},  {\em Proc. SPIE} {\bf 8446-269} (2012).

\bibitem{Schaefer03}
Schaefer, B.~E. et~al., ``{GRB} 021004: a {M}assive {P}rogenitor {S}tar
  {S}urrounded by {S}hells,'' {\em ApJ}~{\bf 588, 387} (2003).

\bibitem{Quimby07}
Quimby, R.~M., Aldering, G., Wheeler, J.~C., H{\"{o}}fflich, P., Akerlof,
  C.~W., and Rykoff, E.~S., ``{SN} 2005ap: {A} {M}ost {B}rilliant
  {E}xplosion,'' {\em ApJ}~{\bf 668, L99} (2007).

\bibitem{Zheng08}
Zheng, C. et~al., ``First-{Y}ear {S}pectroscopy for the {S}loan {D}igital {S}ky
  {S}urvey-{II} {S}upernova {S}urvey,'' {\em AJ}~{\bf 135, 1766} (2008).

\bibitem{Kaspi07}
Kaspi, S., Brandt, W.~N., Maoz, D., Netzer, H., Schneider, D.~P., and Shemmer,
  O., ``Reverberation {M}apping of {H}igh-{L}uminosity {Q}uasars: {F}irst
  {R}esults,'' {\em ApJ}~{\bf 659, 997} (2007).

\bibitem{Corsini08}
Corsini, E.~M., Wegner, G., Saglia, R.~P., Thomas, J., Bender, R., and Thomas,
  D., ``Spatially {R}esolved {S}pectroscopy of {C}oma {C}luster {E}arly-type
  {G}alaxies. {IV}. {C}ompleting the {D}ata {S}et,'' {\em ApJS}~{\bf 175, 462}
  (2008).

\bibitem{Schneider00}
Schneider, D.~P. et~al., ``Discovery of a {P}air of $z = 4.25$ {Q}uasars from
  the {S}loan {D}igital {S}ky {S}urvey,'' {\em AJ}~{\bf 120, 2183} (2000).

\bibitem{Brand03}
Brand, K., Rawlings, S., Hill, G.~J., Lacy, M., Mitchell, E., and Tufts, J.,
  ``Two 100-{M}pc-scale structures in the three-dimensional distribution of
  radio galaxies and their implications,'' {\em MNRAS}~{\bf 344, 283} (2003).

\bibitem{Hornschemeier03}
Hornschemeier, A.~E. et~al., ``The {C}handra {D}eep {F}eild {N}orth {S}urvey.
  {XV}. {O}ptically {B}right, {X}-ray {F}aint {S}ources,'' {\em AJ}~{\bf 126,
  575} (2003).

\bibitem{Verhamme06}
Verhamme, A., Schaerer, D., and Maselli, A., ``I. {U}nderstanding {\lya} line
  profile morphologies,'' {\em A{\&}A}~{\bf 460, 397} (2006).

\bibitem{Barnes11}
Barnes, L.~A., Haehnelt, M.~G., Tescari, E., and Viel, M., ``Galactic winds and
  extended {\lya} emission from the host galaxies of high column density
  quasi-stellar object absorption systems,'' {\em MNRAS}~{\bf 416, 1723}
  (2011).

\bibitem{Kulas12}
Kulas, K.~R., Shapley, A.~E., Kollmeier, J.~A., Zheng, Z., Steidel, C.~C., and
  Hainline, K.~N., ``The {K}inematics of {M}ultiple-peaked {\lya} {E}mission in
  {S}tar-forming {G}alaxies at {$z \sim 2-3$},'' {\em ApJ}~{\bf 745, 33}
  (2012).

\bibitem{Chonis12}
Chonis, T.~S. et~al. {\em In preparation}  (2012).

\bibitem{Adams11}
Adams, J.~J. et~al., ``The {HETDEX} {P}ilot {S}urvey. {I}. {S}urvey {D}esign,
  {P}erformance, and {C}atalog of {E}mission-{L}ine {G}alaxies,'' {\em
  ApJS}~{\bf 192, 5} (2011).

\bibitem{Finkelstein11}
Finkelstein, S.~L. et~al., ``The {HETDEX} {P}ilot {S}urvey. {III}. {T}he {L}ow
  {M}etallicities of {H}igh-{R}edshift {\lya} {G}alaxies,'' {\em ApJ}~{\bf 729,
  140} (2011).

\bibitem{Adams09}
Adams, J.~J., Hill, G.~J., and MacQueen, P.~J., ``B2 0902{$+$}34: a
  {C}ollapsing {P}rotogiant {E}lliptical {G}alaxy at {$z = 3.4$},'' {\em
  ApJ}~{\bf 694, 314} (2009).

\bibitem{McLinden11}
McLinden, E.~M. et~al., ``First {S}pectroscopic {M}easurements of {\oiii}
  {E}mission from {\lya} {S}elected {F}ield {G}alaxies at {$z \sim 3.1$},''
  {\em ApJ}~{\bf 730, 136} (2011).

\bibitem{Chonis11}
Chonis, T.~S., {\em Development of a {N}ew {L}ow {R}esolution {S}pectrograph
  for {P}robing {L}yman-{$\alpha$} {E}mitters in the {HETDEX} {S}urvey},
  Master's thesis, The University of Texas at Austin (2011).

\bibitem{Hill08b}
Hill, G.~J. et~al., ``Design, construction, and performance of {VIRUS-P}: the
  prototype of a highly replicated integral field spectrograph for the {HET},''
  in [{\em Ground-based and Airborne Instrumentation for Astronomy
  II}{\nolinebreak\hspace{0.1em}]},  {\em Proc. SPIE} {\bf 7014-257} (2008).

\bibitem{Dressler11}
Dressler, A. et~al., ``{IMACS}: {T}he {I}namori-{M}agellan {A}real {C}amera and
  {S}pectrograph on {M}agellan-{B}aade,'' {\em PASP}~{\bf 123, 288} (2011).

\bibitem{Sharples12}
Sharples, R.~M. et~al., ``Status of the {KMOS} multi-object near-infrared
  integral field spectrograph,'' in [{\em Ground-based and Airborne
  Instrumentation for Astronomy IV}{\nolinebreak\hspace{0.1em}]},  {\em Proc.
  SPIE} {\bf 8446-18} (2012).

\bibitem{McLean12}
McLean, I.~S., Steidel, C.~C., Epps, H.~W., Matthews, K.~Y., and Adkins, S.~M.,
  ``{MOSFIRE}: the multi-object spectrometer for infrared exploration at the
  {K}eck {O}bservatory,'' in [{\em Ground-based and Airborne Instrumentation
  for Astronomy IV}{\nolinebreak\hspace{0.1em}]},  {\em Proc. SPIE} {\bf
  8446-17} (2012).

\bibitem{Yang10}
Yang, Y., Zabludoff, A., Eisenstein, D., and Dav{\'{e}}, R., ``Strong
  {F}ield-to-field {V}ariation of {\lya} {N}ebulae {P}opulations at {$z \approx
  2.3$},'' {\em ApJ}~{\bf 719, 1654} (2010).

\bibitem{Odewahn12}
Odewahn, S.~C. et~al., ``{VIRUS} {P}arallel {O}bservations with the
  {H}obby-{E}berly {T}elescope,'' in [{\em American Astronomical Society
  Meeting Abstracts}{\nolinebreak\hspace{0.1em}]},  {\em AAS} {\bf 219, 424.18}
  (2012).

\bibitem{Blanc11}
Blanc, G.~A. et~al., ``The {HETDEX} {P}ilot {S}urvey. {II}. the {E}volution of
  the {\lya} {E}scape {F}raction from the {U}ltraviolet {S}lope and
  {L}uminosity {F}unction of {$1.9 < z < 3.8$} {LAE}s,'' {\em ApJ}~{\bf 736,
  31} (2011).

\bibitem{Chonis10}
Chonis, T.~S. et~al., ``Development of a cryogenic system for the {VIRUS} array
  of 150 spectrographs for the {H}obby-{E}berly {T}elescope,'' in [{\em
  Ground-based and Airborne Instrumentation for Astronomy
  III}{\nolinebreak\hspace{0.1em}]},  {\em Proc. SPIE} {\bf 7735-265} (2010).

\bibitem{Murphy08}
Murphy, J.~D. et~al., ``Focal ratio degradation and transmission in {VIRUS-P}
  optical fibers,'' in [{\em Advanced Optical and Mechanical Technologies in
  Telescopes and Instrumentation}{\nolinebreak\hspace{0.1em}]},  {\em Proc.
  SPIE} {\bf 7018, 2} (2008).

\bibitem{AllingtonSmith02}
Allington-Smith, J. et~al., ``Integral {F}ield {S}pectroscopy with the {G}emini
  {M}ultiobject {S}pectrograph. {I}. {D}esign, {C}onstruction, and {T}esting,''
  {\em PASP}~{\bf 114, 892} (2002).

\bibitem{Baldry04}
Baldry, I.~K., Bland-Hawthorn, J., and Robertson, J.~G., ``Volume {P}hase
  {H}olographic {G}ratings: {P}olarization {P}roperties and {D}iffraction
  {E}fficiency,'' {\em PASP}~{\bf 116, 403} (2004).

\bibitem{Burgh07}
Burgh, E.~B., Bershady, M.~A., Westfall, K.~B., and Nordsieck, K.~H.,
  ``Recombination {G}hosts in {L}ittrow {C}onfiguration: {I}mplications for
  {S}pectrographs {U}sing {V}olume {P}hase {H}olographic {G}ratings,'' {\em
  PASP}~{\bf 119, 1069} (2007).

\bibitem{Barden00}
Barden, S.~C., Arns, J.~A., Colburn, W.~S., and Williams, J.~B.,
  ``Volume-{P}hase {H}olographic {G}ratings and the {E}fficiency of {T}hree
  {S}imple {V}olume-{P}hase {H}olographic {G}ratings,'' {\em PASP}~{\bf 112,
  809} (2000).

\bibitem{Gaylord85}
Gaylord, T.~K. and Moharam, M.~G., ``Analysis and applications of optical
  diffraction by gratings,'' {\em Proc. IEEE}~{\bf 73, 894} (1985).

\bibitem{Hill03}
Hill, G.~J., Wolf, M.~J., Tufts, J.~R., and Smith, E.~C., ``Volume {P}hase
  {H}oographic {G}risms for {O}ptical and {I}nfrared {S}pectrographs,'' in
  [{\em Specialized Optical Developments in
  Astronomy}{\nolinebreak\hspace{0.1em}]},  {\em Proc. SPIE} {\bf 4842, 1}
  (2003).

\bibitem{Adams08}
Adams, J.~J., Hill, G.~J., and MacQueen, P.~J., ``Volume phase holographic
  grating performance on the {VIRUS-P} instrument,'' in [{\em Ground-based and
  Airborne Instrumentation for Astronomy II}{\nolinebreak\hspace{0.1em}]},
  {\em Proc. SPIE} {\bf 7014, 232} (2008).

\bibitem{Chonis12b}
Chonis, T.~S., Hill, G.~J., Clemens, J.~C., Dunlap, B., and Lee, H., ``Methods
  for evaluating the performance of volume phase holographic gratings for the
  {VIRUS} spectrograph array,'' in [{\em Ground-based and Airborne
  Instrumentation for Astronomy IV}{\nolinebreak\hspace{0.1em}]},  {\em Proc.
  SPIE} {\bf 8446-209} (2012).

\bibitem{Prochaska12}
Prochaska, T. et~al., ``{HETDEX}: {VIRUS} {S}pectrographs {A}ssembly and
  {A}lignment,'' in [{\em American Astronomical Society Meeting
  Abstracts}{\nolinebreak\hspace{0.1em}]},  {\em AAS} {\bf 219, 424.20} (2012).

\bibitem{Tuttle12}
Tuttle, S.~E. et~al., ``Initial results from {VIRUS} production
  spectrographs,'' in [{\em Ground-based and Airborne Instrumentation for
  Astronomy IV}{\nolinebreak\hspace{0.1em}]},  {\em Proc. SPIE} {\bf 8446-221}
  (2012).

\bibitem{Lee10b}
Lee, H., Hill, G.~J., Marshall, J.~L., Vattiat, B.~L., and Depoy, D.~L.,
  ``Visible {I}ntegral-field {R}eplicable {U}nit {S}pectrograph ({VIRUS})
  optical tolerance,'' in [{\em Ground-based and Airborne Instrumentation for
  Astronomy III}{\nolinebreak\hspace{0.1em}]},  {\em Proc. SPIE} {\bf 7735-140}
  (2010).

\bibitem{Heisler10}
Heisler, J.~T., Good, J.~M., Savage, R.~D., Vattiat, B.~L., Hayes, R.~J.,
  Millison, N.~T., and Soukup, I.~M., ``Integration of {VIRUS} spectrographs
  for the {HET} dark energy experiment,'' in [{\em Ground-based and Airborne
  Telescopes III}{\nolinebreak\hspace{0.1em}]},  {\em Proc. SPIE} {\bf
  7733-153} (2010).

\bibitem{Lee12}
Lee, H. et~al., ``Metrology systems of {H}obby-{E}berly {T}elescope wide field
  upgrade,'' in [{\em Ground-based and Airborne Telescopes
  IV}{\nolinebreak\hspace{0.1em}]},  {\em Proc. SPIE} {\bf 8444-181} (2012).

\bibitem{DES}
{Dark Energy Survey, }
\newblock $<$http://www.darkenergysurvey.org$>$.

\bibitem{Stark10}
Stark, D.~P., Ellis, R.~S., Chiu, K., Ouchi, M., and Bunker, A., ``Keck
  spectroscopy of faint $3<z<7$ {L}yman break galaxies - {I}. {N}ew constraints
  on cosmic reionization from the luminosity and redshift-dependent fraction of
  {L}yman $\alpha$ emission,'' {\em MNRAS}~{\bf 408, 1628} (2010).

\bibitem{Kirkpatrick99}
Kirkpatrick, J.~D. et~al., ``Dwarfs {C}ooler than {M}: {T}he {D}efinition of
  {S}pectral {T}ype {L} {U}sing {D}iscoveries from the 2 {M}icron {A}ll-{S}ky
  {S}urvey ({2MASS}),'' {\em ApJ}~{\bf 519, 802} (1999).

\bibitem{Muzerolle05}
Muzerolle, J., Luhman, K.~L., Briceno, C., Hartmann, L., and Calvet, N.,
  ``Measuring {A}ccretion in {Y}oung {S}ubstellar {O}bjects: {A}pproaching the
  {P}lanetary {M}ass {R}egime,'' {\em ApJ}~{\bf 625, 906} (2005).

\bibitem{Cruz07}
Cruz, K.~L. et~al., ``Meeting the {C}ool {N}eighbors. {IX}. {T}he {L}uminosity
  {F}unction of {M7-L8} {U}ltracool {D}warfs in the {F}ield,'' {\em AJ}~{\bf
  133, 439} (2007).

\bibitem{Luhman09}
Luhman, K.~L. et~al., ``An {I}nfrared/{X}-{R}ay {S}urvey for {N}ew {M}embers of
  the {T}aurus {S}tar-{F}orming {R}egion,'' {\em ApJ}~{\bf 703, 399} (2009).

\end{thebibliography}
\bibliographystyle{spiebib}   

\end{document}